\documentclass[nofootinbib,aps,11pt,amsmath,amssymb,a4paper,prd,preprintnumbers]{revtex4-1}

\usepackage[american]{babel}
\usepackage[utf8]{inputenc}
\usepackage[T1]{fontenc}

\usepackage{graphicx}
\usepackage{xcolor}

\usepackage[caption=false]{subfig}

\usepackage{slashed}

\usepackage{units}

\allowdisplaybreaks

\usepackage{hyperref}
\hypersetup{
   pdfnewwindow=true,    
   colorlinks=true,      
   linkcolor=blue,       
   citecolor=blue,       
   filecolor=black,       
   urlcolor=blue         
}

\begin{document}

\title{\Large {\bf{Baryonic Higgs at the LHC}}}
\author{Michael Duerr,$^{1}$ Pavel Fileviez P\'erez,$^{2}$ Juri Smirnov$^{3}$}        
\affiliation{\vspace{3ex} $^{1}$DESY, Notkestra\ss{}e 85, D-22607 Hamburg, Germany \\
$^{2}$Physics Department, Center for Education and Research in Cosmology and Astrophysics (CERCA), and Institute for the Science of Origins (ISO), Case Western Reserve University, Rockefeller Bldg. 2076 Adelbert Rd. Cleveland, OH 44106, USA\\
$^{3}$INFN, Sezione di Firenze, and Department of Physics and Astronomy, University of Florence, Via G.\ Sansone 1, 50019 Sesto Fiorentino, Italy} 
\email[E-mail addresses: ]{michael.duerr@desy.de}
\email{pxf112@case.edu}
\email{juri.smirnov@fi.infn.it}
%
\begin{abstract}    
We investigate the possible collider signatures of a new Higgs in simple extensions of the Standard Model where baryon number is a local symmetry spontaneously 
broken at the low scale. We refer to this new Higgs as ``Baryonic Higgs''. This Higgs has peculiar properties since it can decay into all Standard Model particles, the leptophobic gauge boson, 
and the vector-like quarks present in these theories to ensure anomaly cancellation. We investigate in detail the constraints from the $\gamma \gamma$, $Z\gamma$, $ZZ$, and $WW$ 
searches at the Large Hadron Collider, needed to find a lower bound on the scale at which baryon number is spontaneously broken. The di-photon channel 
turns out to be a very sensitive probe in the case of small scalar mixing and can severely constrain the baryonic scale.
We also study the properties of the leptophobic gauge boson in order to understand the testability of these theories at the LHC. 
\end{abstract}

\preprint{DESY 17-050}

\maketitle

\section{Introduction}
\label{sec:Introduction}
The discovery of the Standard Model~(SM) spin-zero boson, the Brout--Englert--Higgs boson, at the Large Hadron Collider~(LHC) was crucial to establish the mechanism of electroweak symmetry breaking. Thanks to this great discovery, we know that the electroweak symmetry, 
$SU(2)_L \otimes U(1)_Y$, is broken spontaneously  and all masses for the elementary particles are proportional to only one symmetry breaking scale, $v_0=\unit[246]{GeV}$.
Now, the primary goal of the LHC experiments is to look for new physics which could give rise to exotic signatures. There are many appealing extensions of the Standard Model and most of them contain a new Higgs sector which can give rise to new signatures at collider experiments. 
See for example Ref.~\cite{Olive:2016xmw} for the review of the Higgs sector of some extensions of the Standard Model. 

In this article we investigate the properties of the new 
physical Higgs in extensions of the Standard Model where the global baryon number of the Standard Model is promoted to a local symmetry. The idea of having baryon 
number as a local symmetry was mentioned in Refs.~\cite{Lee:1955vk,Pais:1973mi}; see also Refs.~\cite{Rajpoot:1987yg,Foot:1989ts,Carone:1995pu} for some theoretical considerations.
Recently, this idea has been investigated in detail and a class of realistic theories has been proposed~\cite{FileviezPerez:2010gw,FileviezPerez:2011pt,Duerr:2013dza,Perez:2014qfa,Perez:2015rza}. A simple UV completion of these theories has been proposed 
in Ref.~\cite{FileviezPerez:2016laj}, see also Ref.~\cite{Fornal:2015boa}. These theories have the following features:
\begin{itemize}
\item In this context one can understand the spontaneous breaking of the local baryon number at the low scale.
\item Some of these theories predict that the proton is stable or very long-lived. Then, one can think about the possibility to have unification of gauge interactions at the low scale. 
\item These theories predict the existence of vector-like quarks which cancel all baryonic anomalies.
\item A leptophobic gauge boson is always present in these theories.
\item Some of these theories predict the existence of a dark matter candidate and one can make a connection between the baryon and dark matter asymmetries.
\end{itemize}
See Refs.~\cite{Dulaney:2010dj,Duerr:2013lka,Perez:2013tea,Duerr:2014wra,Ohmer:2015lxa,Perez:2014kfa,Duerr:2015vna} for the study of some phenomenological and cosmological aspects of these theories.

In this article we investigate the possible signatures at the LHC from the decays of the new Higgs in these theories. We structure our discussion as follows. In Section~\ref{sec:LocalBaryonNumber} we present general features of these theories and discuss the scalar sector in detail. In Section~\ref{sec:BaryonicHiggsLeptophobicGaugeBoson} we discuss production and decay of the baryonic Higgs and show that non-trivial constraints are coming from $\gamma \gamma$, $Z\gamma$, $ZZ$, and $WW$ searches. We find scenarios in which striking signatures at the LHC are expected since the baryonic Higgs branching ratio 
into two photons can be large compared to the branching ratio of the SM Higgs into two photons. To complete the discussion, we also study the properties of the leptophobic gauge boson. Our results motivate the search for these signatures at the LHC which are crucial to test these theories. We summarize our results in Section~\ref{sec:Summary}.

\section{Simple Theories  for Baryon Number}
\label{sec:LocalBaryonNumber}
\subsection{Theoretical Framework}
\label{sec:Setup}

It is well known that in the Standard Model baryon number is an accidental global symmetry which is only conserved at the classical level and broken 
at the quantum level by the $SU(2)$-instantons. One can think about defining a simple extension of the Standard Model where baryon number is a local symmetry~\cite{Pais:1973mi,Foot:1989ts,Carone:1995pu}. In such a scenario one can study the spontaneous breaking of baryon number and investigate the possible implications for particle physics and cosmology. Realistic versions of these theories have been 
proposed~\cite{FileviezPerez:2010gw,FileviezPerez:2011pt,Duerr:2013dza,Perez:2014qfa,Perez:2015rza} and various phenomenological aspects have been studied~\cite{Dulaney:2010dj,Duerr:2013lka,Perez:2013tea,Duerr:2014wra,Ohmer:2015lxa,Perez:2014kfa,Duerr:2015vna}. These theories are based on the gauge group
\begin{equation}
G_B = SU(3)_C \otimes SU(2)_L \otimes U(1)_Y \otimes U(1)_B .
\end{equation}

Recently, a simple UV completion of these theories has been proposed in Ref.~\cite{FileviezPerez:2016laj}. In this context one always predicts the existence of vector-like quarks to define an anomaly-free theory. Using this result as a motivation we will investigate theories for baryon number where the baryonic anomalies are cancelled via the introduction of vector-like quarks with particular baryon numbers that can differ from the baryon number of the SM quarks.

\begin{table}[t]
 \caption{Particle content. 
 \label{tab:particles}}
 \begin{tabular}{ccccc}
 \hline\hline
 Field & $SU(3)_C$ & $SU(2)_L$ & $U(1)_Y$ & $U(1)_B$ \\
 \hline
 Fermions & & & & \\
 $\ell_L$ & $1$ & $2$ & $-1/2$ & $0$ \\
  $e_R$ & $1$ & $1$ & $-1$ & $0$ \\
 $q_L$ & $3$ & $2$ & $1/6$ & $1/3$ \\
 $u_R$ & $3$ & $1$ & $2/3$ & $1/3$ \\
 $d_R$ & $3$ & $1$ & $-1/3$ & $1/3$ \\
 ${Q}_L$ & $3$ & $2$ & $Y_1$ & $B_1$ \\
 ${Q}_R$ & $3$ & $2$ & $Y_1$ & $B_2$ \\
 $U_R$ & $3$ & $1$ & $Y_2$ & $B_1$ \\
 $U_L$ & $3$ & $1$ & $Y_2$ & $B_2$ \\
 $D_R$ & $3$ & $1$ & $Y_3$ & $B_1$ \\
 $D_L$ & $3$ & $1$ & $Y_3$ & $B_2$ \\
 \hline
 Scalars & & & & \\
 $S_B$ & $1$ & $1$ & $0$ & $B_2 - B_1$ \\
 $H$ & $1$ & $2$ & $1/2$ & $0$ \\
 \hline \hline
 \end{tabular}
\end{table}

In Table~\ref{tab:particles} we list the quantum numbers of the SM particles and the new vector-like quarks introduced to cancel the anomalies. All relevant anomalies are cancelled if 
one requires~\cite{FileviezPerez:2011pt,Duerr:2013lka}
\begin{equation}
 B_1-B_2=- \frac{1}{n_f}
\end{equation}
for the baryon numbers and 
\begin{equation}\label{eq:hyperchargeSolution}
Y_2=Y_1 \mp \frac{1}{2}  \text{ and } Y_3= Y_1 \pm \frac{1}{2}
\end{equation}
for the hypercharges of the vector-like quarks. Here, $n_f$ is the number of copies of the new vector-like quarks. We will assume $n_f=3$ in the remainder of this article since in UV completions of these theories, 
such as the one proposed in Ref.~\cite{FileviezPerez:2016laj}, one needs three copies of vector-like quarks.

There are many possible solutions to Eq.~\eqref{eq:hyperchargeSolution}. In this article, we show the numerical results only for three simple scenarios where at least one of the new quarks 
has the same hypercharge as one of the SM quarks. The scenarios we consider are:\footnote{See Ref.~\cite{Duerr:2016eme} for a study of the same solutions in a different context.}
\begin{itemize}
\item Scenario I: $Y_1=1/6$, $Y_2=2/3$ and $Y_3=-1/3$.
\item Scenario II: $Y_1=-5/6$, $Y_2=-4/3$ and $Y_3=-1/3$.
\item Scenario III: $Y_1=7/6$, $Y_2=5/3$ and $Y_3=2/3$.   
\end{itemize}

The vector-like quarks must be heavy to be in agreement with experimental bounds and their masses are generated after spontaneous breaking of baryon number. The relevant interactions are
\begin{align}
\label{eq:Lagrangian}
- \mathcal{L}_\text{VLQs} &= h_1 \overline{Q}_L \tilde{H} U_R + h_2 \overline{Q}_L H D_R + h_3 \overline{Q}_R \tilde{H} U_L + h_4 \overline{Q}_R H D_L \nonumber \\
& \quad + \lambda_Q \overline{Q}_R Q_L S_B + \lambda_U \overline{U}_L U_R S_B + \lambda_D \overline{D}_L D_R S_B \ + \  \text{h.c.},
\end{align}
where $S_B$ is the new Higgs boson responsible for the breaking of baryon number, see Table~\ref{tab:particles} for its quantum numbers. 
Notice that for baryon numbers $B_i$ different from $1/3$, there is no mixing between the SM quarks and the new vector-like quarks, 
and there are no new sources of flavor violation.  It is important to emphasize that in general the new quarks do not mix with the SM quarks unless 
one chooses very specific values for their baryon numbers and hypercharges. 

Let us now list some of the main predictions of these models:
\begin{itemize}
\item One predicts the existence of a leptophobic gauge boson $Z_B$, and the local baryon number can be broken at the low scale in agreement with all experimental constraints.
\item Since we stick to the case $n_f=3$, the baryon number of the baryonic Higgs $S_B$ is fixed to $1/3$, and one will generate dimension-9 operators for proton decay such as 
\begin{equation}
 \mathcal{O}_9 = \frac{c_9 \left( S_B^\dagger \right)^3}{\Lambda^5} (u_R u_R d_R e_R).
\end{equation}
For $c_9 \sim 1$ and the vev of $S_B$ around a TeV, one needs $\Lambda > \unit[10^4]{TeV}$ to satisfy the proton decay bounds. Therefore, proton decay is suppressed even if the cutoff of the theory is not very large.
\item In the minimal version of these models the lightest new quark could be stable and can give rise to very exotic signatures at the LHC. 
The vector-like quarks are produced in pairs through QCD processes and could hadronize forming exotic neutral and charged bound states. These bound states 
can give rise to very exotic signatures in the electromagnetic and hadronic calorimeters. In order to avoid all cosmological constraints one can assume 
that the reheating temperature is much smaller than the mass of the lightest new vector-like quark. See for example Ref.~\cite{BBN} for a review on 
the cosmological bounds for long-lived charged fields. We will investigate these signatures in a 
future publication but let us discuss this issue in more details. One can imagine several scenarios for the new quarks:

\begin{itemize}

\item If the new quarks have the same hypercharge as the SM quarks and $B_1$ or $B_2$ is equal to 1/3, the new quarks can mix with the SM quarks and decay due to a bare mass term in the Lagrangian.

\item When the new heavy quarks have different hypercharges and baryon numbers from the SM values they do not mix with the SM quarks and can form bound states between themselves. 
In this case the lightest bounded state is a neutral heavy pion which also decays into two photons. The stable heavy baryons are not abundant as their mass exceeds the rehearing temperature.

\item If one of the new heavy quarks has SM hypercharge and the interaction between the heavy quark, SM quark and the new Higgs $S_B$ is allowed after symmetry breaking the heavy quarks will mix with the SM quarks and decay. For example the term $\overline{Q}_R q_L S_B$ is allowed if $B_2=2/3$ and $Y_1=1/6$.

\item One can also imagine the case proposed in Ref.~\cite{FileviezPerez:2011pt} where one adds 
a new Higgs $X \sim (1,1,0,1/3 - B_2)$ which does not acquire the vacuum expectation value. 
In this case the $X$ field is a cold dark matter candidate and the new quarks decay into the SM quarks and dark matter.
See Ref.~\cite{FileviezPerez:2011pt} for the details.

\end{itemize}
\end{itemize}
%
\subsection{Scalar Sector}
\label{sec:ScalarSector}
In this theory one has two Higgs fields, the SM Higgs $H$ and the baryonic Higgs $S_B$ breaking local baryon number $U(1)_B$. 
Therefore, the scalar potential is given by 
\begin{align}
V \left(H, S_B\right) & =  \mu_H^2 H^\dagger H  + \mu_B^2 S_B^\dagger S_B + \lambda_{HB} \left( H^\dagger H \right) \left( S_B^\dagger S_B\right) 
+ \lambda_H \left( H^\dagger H\right)^2 +\lambda_B \left( S_B^\dagger S_B\right)^2.
\end{align}
In the unitary gauge we can write the SM Higgs multiplet as
\begin{align}
 H   &= \frac{1}{\sqrt{2}} \left( 0, \ h + v_0 \right)^T
\end{align}
and the new spin-zero boson as
\begin{align}
 S_B &= \frac{1}{\sqrt{2}} (s_B + v_B + i A_B) ,
\end{align}
where $v_0 = \unit[246]{GeV}$ is the vacuum expectation value~(vev) of the SM Higgs and $v_B$ is the baryonic symmetry breaking scale. $A_B$ is unphysical and will be eaten by the $Z_B$.
After symmetry breaking, there is mixing between the SM Higgs and the baryonic Higgs, and the mixing angle is given by
\begin{equation}\label{eq:mixingAngle}
 \tan 2 \theta_B \equiv \frac{\lambda_{HB} v_0 v_B}{\lambda_H v_0^2 - \lambda_B v_B^2}.
\end{equation}
The physical Higgs fields can be defined as
\begin{align}
 h_{1} &= h \cos \theta_B + s_B \sin \theta_B, \\
 h_B &= s_B \cos \theta_B - h \sin \theta_B.
\end{align}
The masses of $h_1$ and $h_B$ are given by
\begin{align}
 M_{h_1}^2 &= \lambda_H v_0^2 + \lambda_B v_B^2 + \left( \lambda_H v_0^2 - \lambda_B v_B^2 \right) \sqrt{1 + \frac{v_0^2 v_B^2 \lambda_{HB}^2}{\left(\lambda_H v_0^2 - \lambda_B v_B^2\right)^2}} ,\label{eq:physicalHiggsMasses1} \\
 M_{h_B}^2 &= \lambda_H v_0^2 + \lambda_B v_B^2 - \left( \lambda_H v_0^2 - \lambda_B v_B^2 \right) \sqrt{1 + \frac{v_0^2 v_B^2 \lambda_{HB}^2}{\left(\lambda_H v_0^2 - \lambda_B v_B^2\right)^2}} .\label{eq:physicalHiggsMasses2} 
\end{align}
Here $h_1$ is the mostly SM-like Higgs state with mass $M_{h_1} = \unit[125]{GeV}$. The expressions for the masses given in Eqs.~\eqref{eq:physicalHiggsMasses1} and \eqref{eq:physicalHiggsMasses2} allow for $M_{h_1}$ being larger or smaller than $M_{h_B}$, in agreement with the sign of the mixing angle in Eq.~\eqref{eq:mixingAngle}.\footnote{This corrects an issue with the sign of the mixing angle present in the formulas given in Ref.~\cite{Duerr:2014wra}. This was already noted and corrected in Ref.~\cite{Ohmer:2015lxa}.} The parameters $\lambda_H$, $\lambda_B$, and $\lambda_{HB}$ can be expressed in terms of the other free parameters as
\begin{align}
 \lambda_H    &= \frac{1}{4 v_0^2} \left[ M_{h_1}^2 + M_{h_B}^2 + \left( M_{h_1}^2 - M_{h_B}^2\right) \cos 2 \theta_B \right], \\
 \lambda_B    &= \frac{1}{4 v_B^2} \left[ M_{h_1}^2 + M_{h_B}^2 + \left( M_{h_B}^2 - M_{h_1}^2\right) \cos 2 \theta_B \right] , \\
 \lambda_{HB} &= \frac{1}{2 v_0 v_B} \left( M_{h_1}^2 - M_{h_B}^2 \right) \sin 2 \theta_B.
\end{align}
See the appendix of Ref.~\cite{Duerr:2016tmh} for a similar discussion. Then, in the scalar sector the model has only two free parameters,
\begin{equation}
 M_{h_B} \ \text{and} \  \theta_B. \nonumber
\end{equation}
Note that the vev $v_B$ is related to the mass of the leptophobic gauge boson, which is given by
\begin{equation}
 M_{Z_B} = \frac{1}{n_f} g_B v_B,
\end{equation}
where $g_B$ is the gauge coupling of $U(1)_B$. 

The mixing angle between the SM Higgs and the baryonic Higgs is constrained by the SM Higgs signal strength, whose current value is given by~\cite{Khachatryan:2016vau}
\begin{equation}
 \mu = 1.09 \pm 0.11 .
\end{equation}
At 95\% CL, this leads to a bound of 
\begin{equation}
 \theta_B \leq 0.36 .
\end{equation}
Knowing the features of the model we are ready to investigate the phenomenological properties of the Higgs sector in the next section.

\section{Baryonic Higgs and Leptophobic Gauge Boson}
\label{sec:BaryonicHiggsLeptophobicGaugeBoson}

\subsection{Higgs Decays}
\label{sec:BaryonicHiggsDecays}

\begin{figure}[t]
\includegraphics[width=0.45\linewidth]{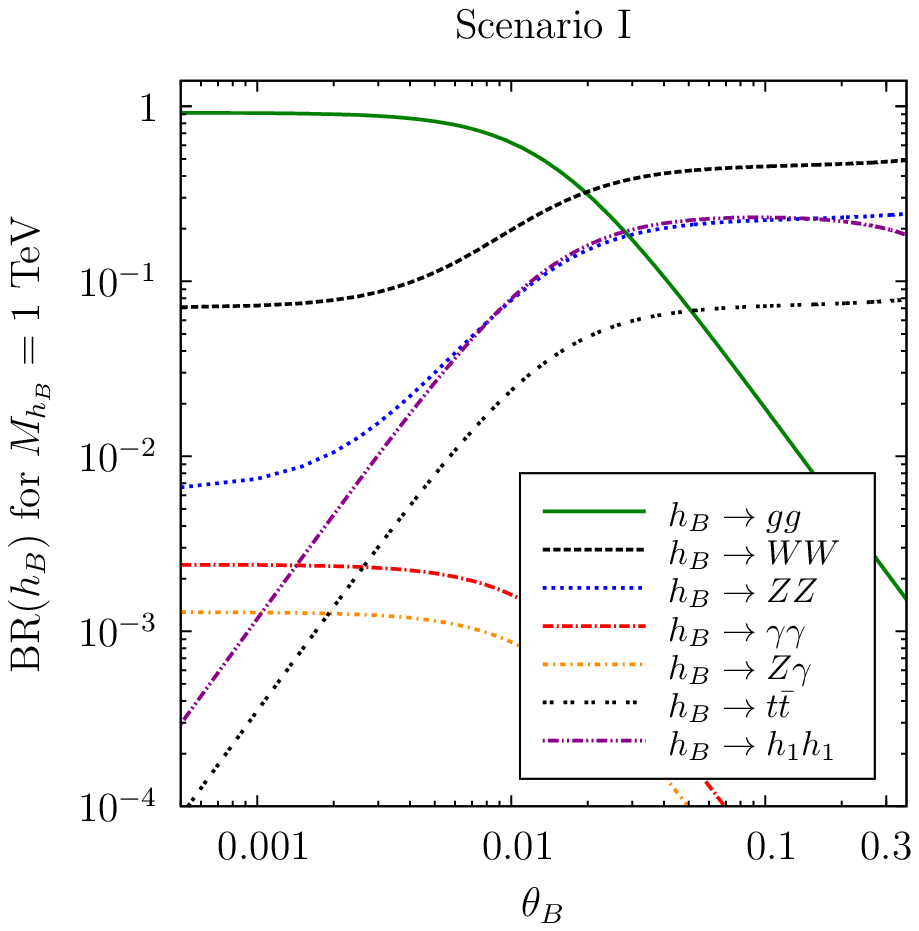}\hspace{.07\linewidth}
\includegraphics[width=0.45\linewidth]{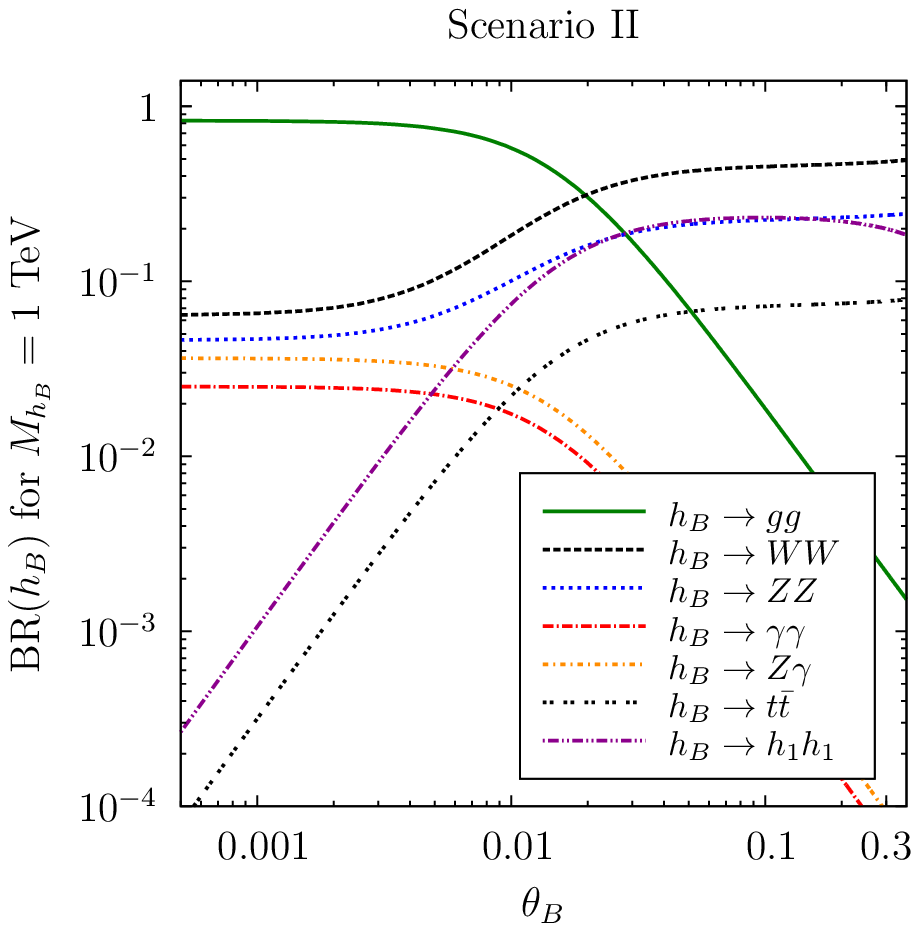}
\includegraphics[width=0.45\linewidth]{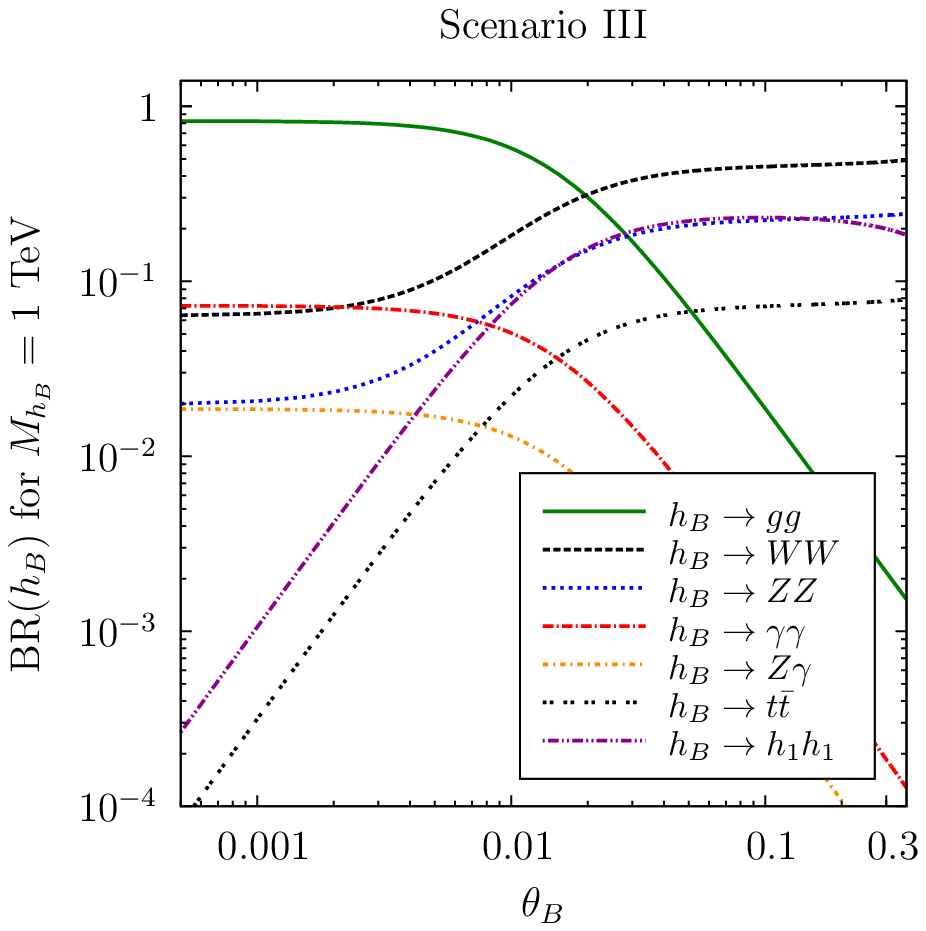}
\caption{Branching ratios of the baryonic Higgs as a function of the mixing angle $\theta_B$ assuming only decays into SM particles. The vector-like quarks are assumed to be heavy and we use $v_B =  \unit[2]{TeV}$ and $M_{h_B} = \unit[1]{TeV}$. 
\label{fig:Brvstheta}}
\end{figure}

The new Higgs $h_B$ can decay into all particles present in the model, i.e.\
$$h_B \to \bar{e}_i e_i, \bar{q}_i q_i, WW, ZZ, h_1 h_1, gg, \gamma \gamma, \bar{U}_k U_k, \bar{D}_k D_k, Z_B Z_B,$$
where $i=1,2,3$ and $k=1,\ldots,6$. The properties of the decays depend in a significant way on the mixing angle between the SM Higgs and the new Higgs. 

In Figure~\ref{fig:Brvstheta} we show the branching ratios of the baryonic Higgs into SM fields as a function of the Higgs mixing angle in the three scenarios. 
We use $M_{h_B}=\unit[1]{TeV}$ for illustration purposes. For large mixing angles close to the upper bound, the tree-level decays via mixing with the SM Higgs dominate, while 
for small mixing angles the loop-induced decays take over. In addition to the decay channels shown, tree-level decays into the leptophobic gauge boson are possible if the new Higgs is heavy enough. 
Depending on the mass hierarchy, it can also decay to some or all of the new quarks. For Figure~\ref{fig:Brvstheta} we have assumed that the new quarks are heavy so that these decays 
are not allowed. One can appreciate that for $\theta_B < 0.02$ one has a large branching ratio into two gluons, while for $\theta_B > 0.02$ the decays into $WW$ dominate. 

\begin{figure}[t]
\includegraphics[width=0.45\linewidth]{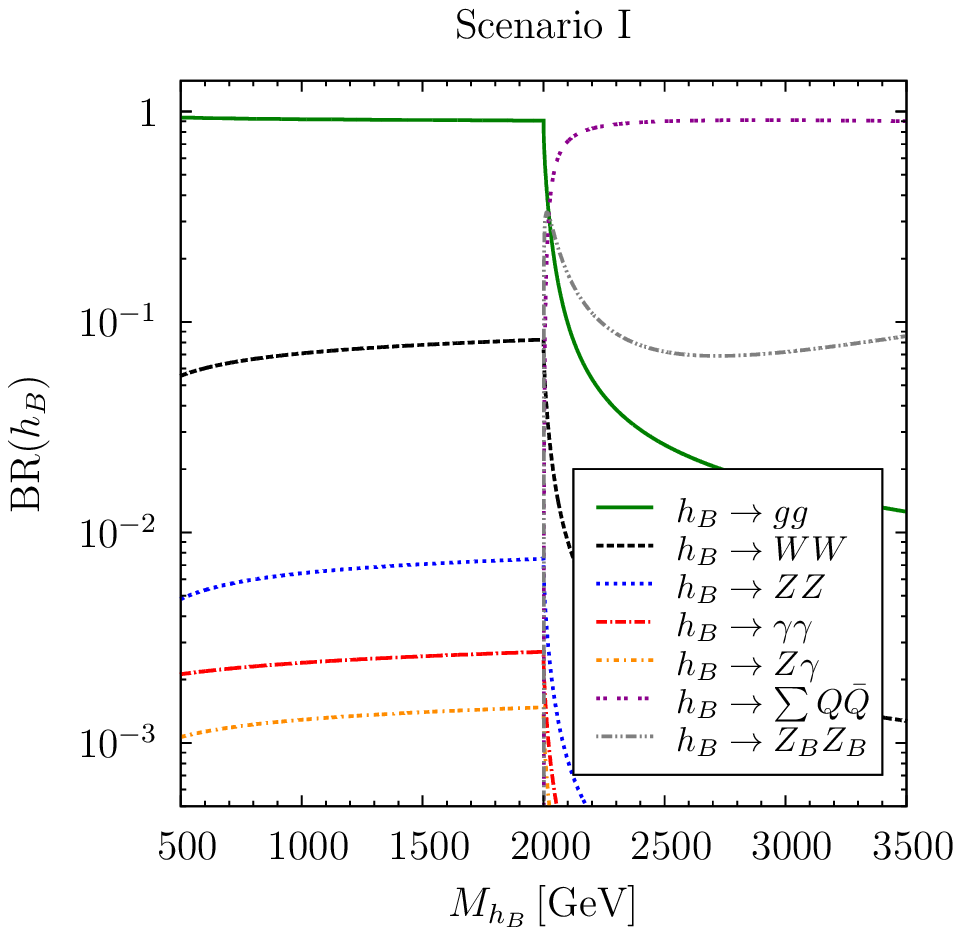}\hspace{.07\linewidth}
\includegraphics[width=0.45\linewidth]{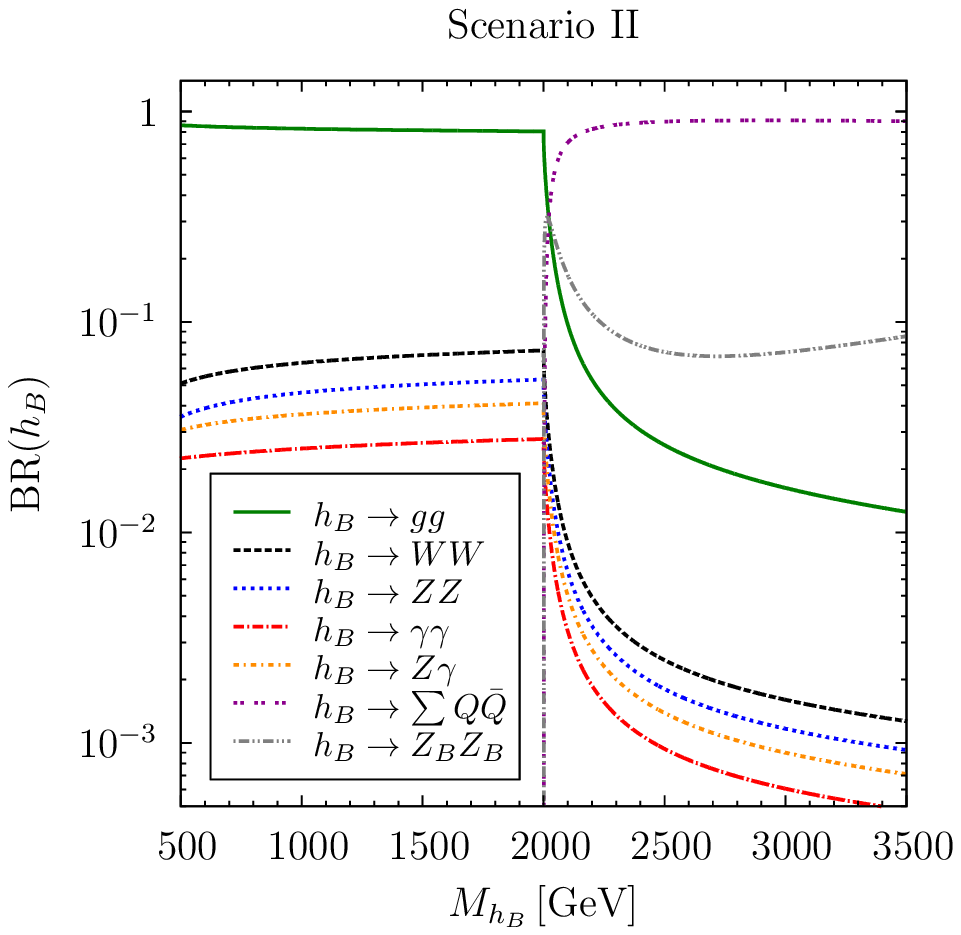}
\includegraphics[width=0.45\linewidth]{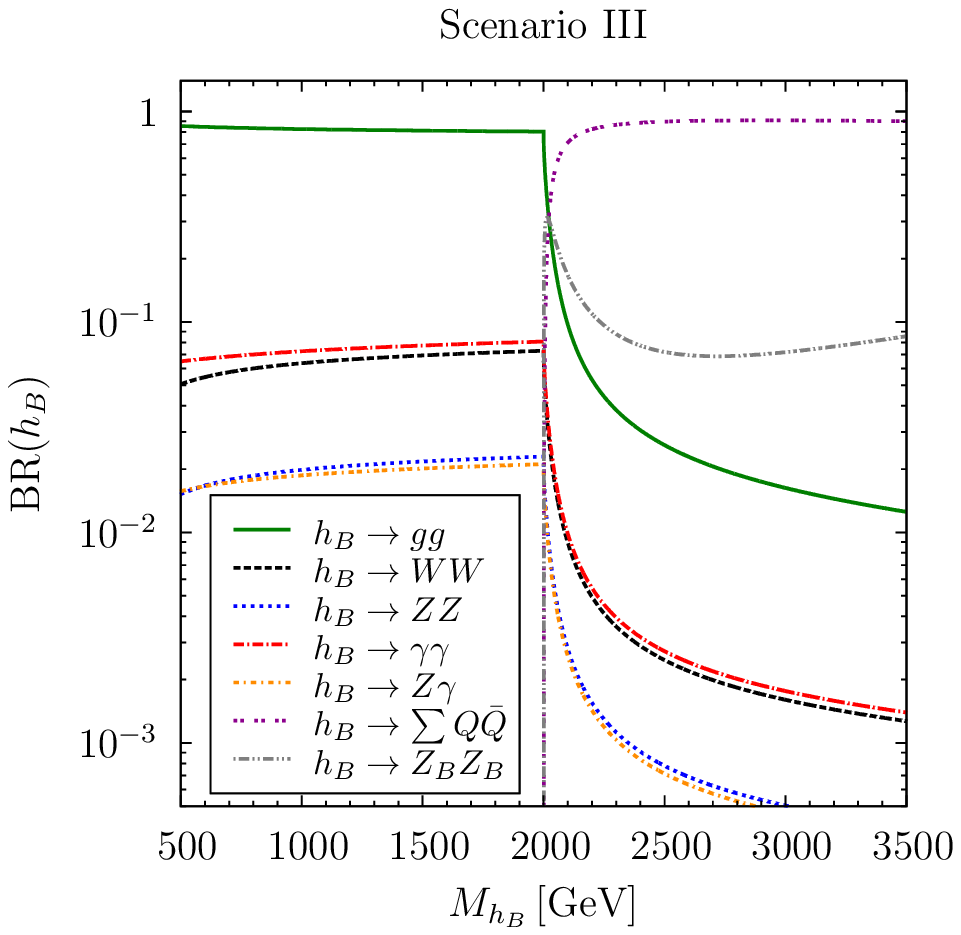}
\caption{Branching ratios of the baryonic Higgs as a function of its mass $M_{h_B}$ assuming negligible mixing angle $\theta_B$. We use $M_{Z_B} = \unit[1]{TeV}$, $v_B  = \unit[20]{TeV}$ and $m_Q = \unit[1]{TeV}$. 
\label{fig:Brs-mhB}}
\end{figure}

For a negligible mixing angle between the SM Higgs and the baryonic Higgs, the branching ratios of the baryonic Higgs are shown in Figure~\ref{fig:Brs-mhB} for the three scenarios and all possible decay modes. 
One can see that in all scenarios the decays into gluons dominate in the range where the decays into vector-like quarks are not allowed, and for $M_{h_B} > 2 m_Q$ the 
decays in two vector-like quarks are dominant. For simplicity we assume throughout the article that all new vector-like quarks have the same mass $m_Q$. Notice that in the second and third scenario the decays into two photons can have a large branching ratio, much larger than $\text{BR}(h_1 \to \gamma \gamma)$ in the SM. Also the branching ratio $\text{BR}(h_B \to Z \gamma)$ is large in scenarios II and III. These results are crucial to understand the constraints from the different search channels that we will discuss later.  
%
\subsection{Baryonic Higgs Production Mechanisms at the LHC} 
\label{sec:BaryonicHiggsProduction}
The new Higgs $h_B$ can be produced through different mechanisms. We focus on the production channels which do not rely on a large mixing angle and can be used to test these models in a generic way. The $h_B$ can be produced through gluon fusion, $gg \to h_B$, through vector-boson fusion via the $Z_B$, $q \bar{q} \to q \bar{q} h_B$, and one can have the associated production 
of $h_B$ and the leptophobic gauge boson $Z_B$, $pp \to Z_B^* \to Z_B h_B$. The associated production with two new quarks is also possible. 

Since the masses of the vector-like quarks are generated once $S_B$ acquires a vev, the couplings between $h_B$ and the new quarks 
are typically large. Therefore, one can produce $h_B$ through gluon fusion where the vector-like quarks are running in the loop and this is the dominant production mechanism at the LHC. The production cross section for this channel is given by 
 \begin{equation}
  \sigma \left( p p \to h_B \right) = \frac{C_{gg}}{s M_{h_B} } \Gamma \left( h_B \to g g \right) .
 \end{equation}
 Here, $C_{gg}$ is the gluonic PDF contribution, and we use the MSTW2008NLO PDFs~\cite{Martin:2009iq} in the remainder of this article, and $s$ is the center-of-mass energy squared. 
 As we have discussed above, the branching ratio into two gauge bosons can be large. In particular, in scenarios II and III we expect a large number of events with two photons in the final state.
 
\subsection{Experimental Constraints and Signatures} 
\label{sec:BaryonicHiggsConstraints}

The cross section for the new Higgs being produced through gluon fusion and decaying into two SM gauge bosons $V$ is given by
\begin{align}
& \sigma_{VV} \equiv \sigma (pp \to h_B) \times \text{BR}(h_B \to V\,V) 
\approx  \frac{C_{gg}}{ s} \frac{2 \, \alpha_s^2 \, M_{h_B}^2 n_f^2 }{9 \pi^3 v_B^2}\text{BR} (h_B \to V\,V), 
\label{eq:GeneralConstraint}
\end{align}
where $n_f = 3$ is the number of generations of vector-like quarks. This expression is general for the case that the new vector-like quark generation lives in the fundamental representation of $SU(3)_C$ and is heavy compared to the baryonic Higgs. 
We can use Eq.~\eqref{eq:GeneralConstraint} to place a limit on the branching ratio of the new Higgs into two gauge bosons using the current LHC results from diboson searches. 

\begin{figure}[tb]
\includegraphics[width=0.45\linewidth]{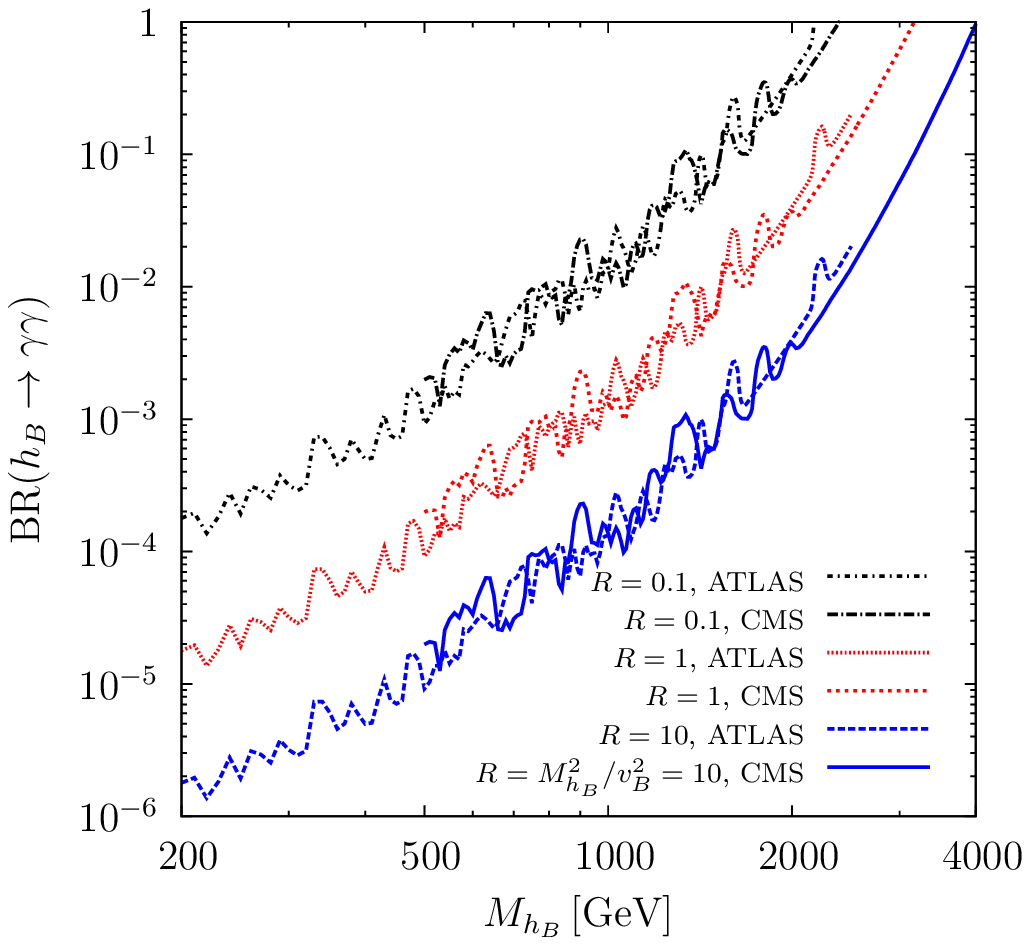}\hspace{.07\linewidth}
\includegraphics[width=0.45\linewidth]{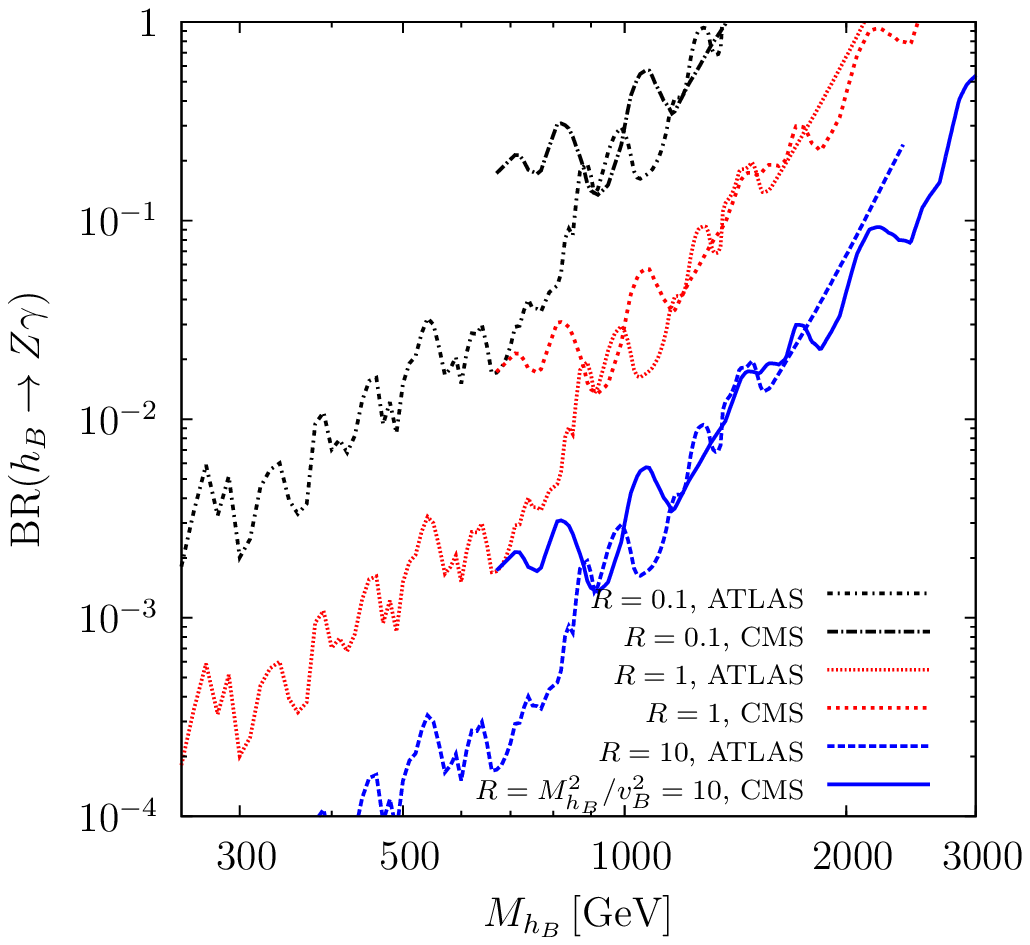}
\vspace{2ex}
\includegraphics[width=0.45\linewidth]{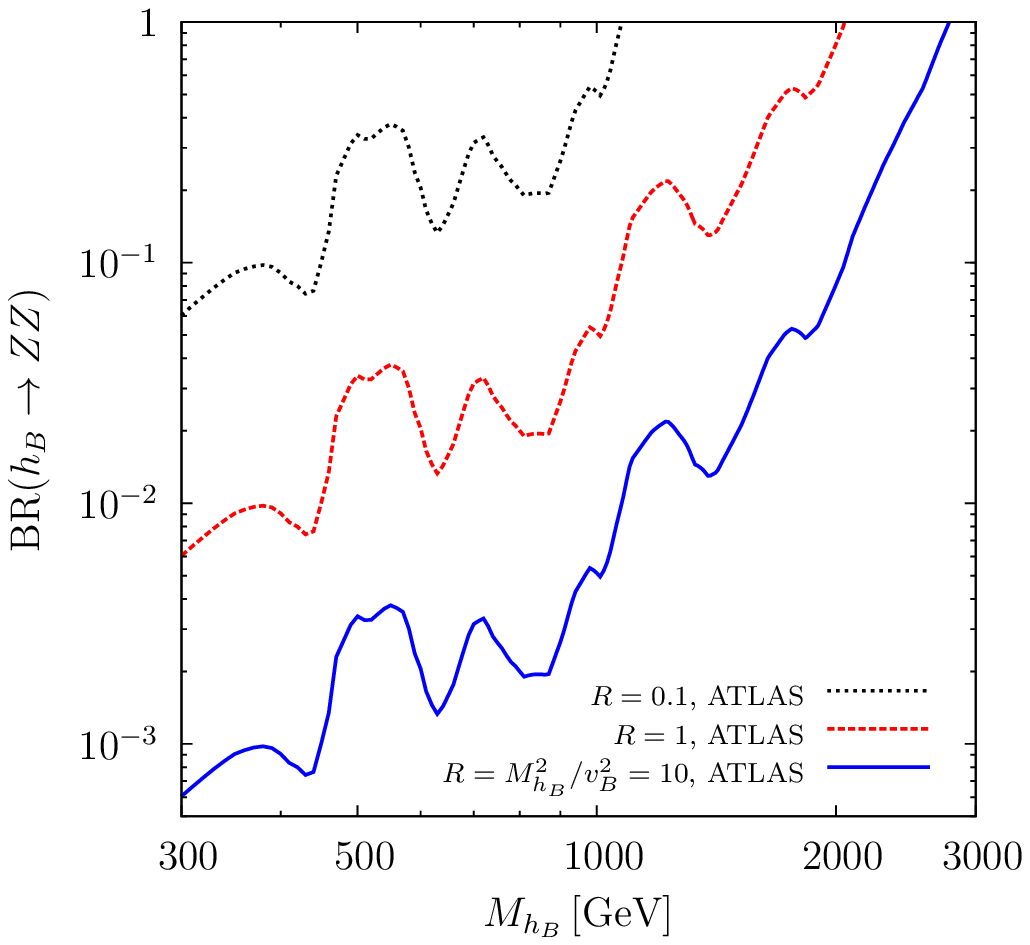}\hspace{.07\linewidth}
\includegraphics[width=0.45\linewidth]{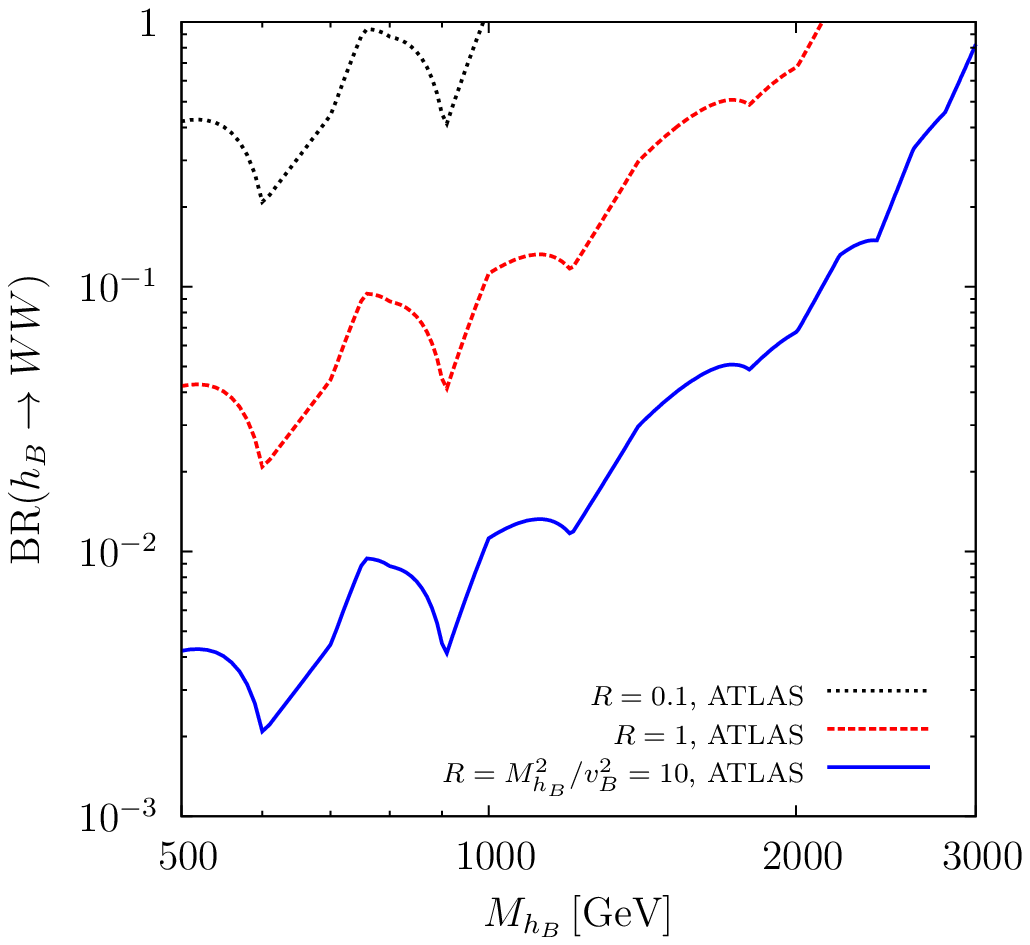}
\caption{Constraints on the branching ratios of $h_B$ into two photons~\cite{ATLAS:2016eeo,Khachatryan:2016yec}, $Z\gamma$~\cite{CMS:2016cbb,ATLAS:2016lri}, $ZZ$~\cite{ATLAS:2016npe}, and $WW$~\cite{ATLAS:2016cwq} for different choices of the ratio between mass and vev of the baryonic Higgs field. In the plots $n_f = 3$.
\label{fig:GeneralLimitDiphoton}}
\end{figure}

In Figure~\ref{fig:GeneralLimitDiphoton} we show the upper limits on the branching ratios for the $\gamma \gamma$~\cite{ATLAS:2016eeo,Khachatryan:2016yec}, $Z \gamma$~\cite{CMS:2016cbb,ATLAS:2016lri}, $ZZ$~\cite{ATLAS:2016npe}, and $WW$~\cite{ATLAS:2016cwq} channels. Notice that these bounds are valid for all scenarios since the new vector-like quarks live in the fundamental representation of the QCD gauge group. These bounds are non-trivial even when we change 
the ratio $R=M_{h_B}^2/v_B^2$ between 0.1 and 10.

\begin{figure}[tb]
\includegraphics[width=0.48\linewidth]{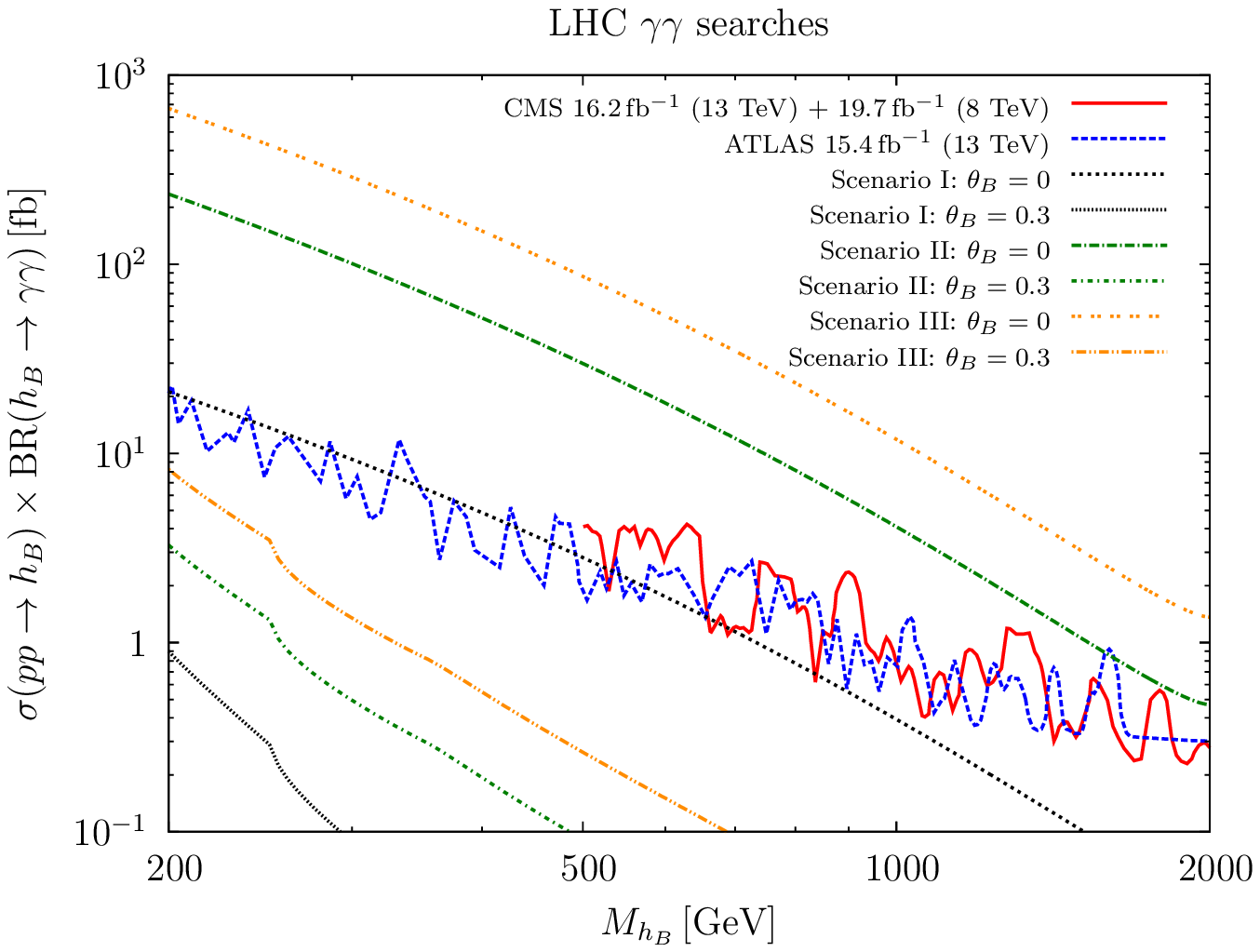}\hspace{.03\linewidth}
\includegraphics[width=0.48\linewidth]{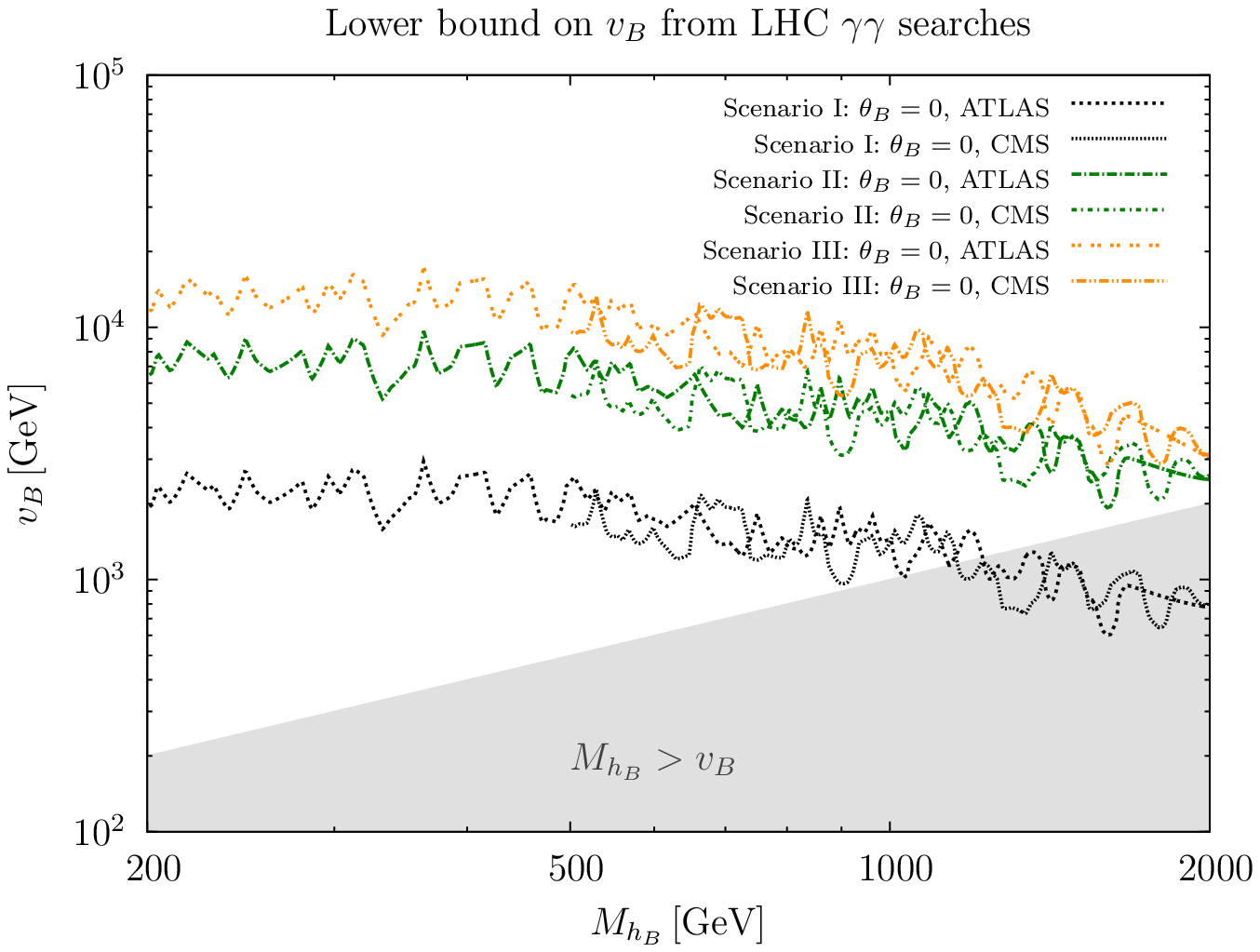}
\caption{Constraints from diphoton searches~\cite{ATLAS:2016eeo,Khachatryan:2016yec}.
Left panel: experimental bounds as well as cross section predictions in scenarios I, II, and III as a function of the baryonic Higgs mass. Cross sections are shown both for negligible mixing angle and for a near maximal value of $\theta_B = 0.3$. 
Right panel: experimental bounds translated into lower limits on $v_B$ as a function of $M_{h_B}$ for the case of negligible mixing. In the left panel we use $v_B=\unit[2]{TeV}$, and in both panels $n_f=3$ and $m_Q=\unit[1]{TeV}$.
\label{fig:diphotonLimits}}
\end{figure}

Let us now discuss the different channels, starting with the di-photon searches. Assuming that the vector-like quarks are heavier than the baryonic Higgs, it is a good approximation to write (for one fermion) 
\begin{equation}
\Gamma (h_B \to \gamma \gamma) = \frac{9 \alpha^2 n_f^2 \,Q^4\,M_{h_B}^3}{144 \pi^3\,v_B^2} 
\end{equation}
for the decay width into two photons. Here, $Q$ is the electric charge of a vector-like quark running in the loop. In the left panel of Figure~\ref{fig:diphotonLimits}, we show the most current limits 
on the production of a resonance and decay into two photons from ATLAS~\cite{ATLAS:2016eeo} and CMS~\cite{Khachatryan:2016yec}. 
We also show the predicted cross sections in the three scenarios for a particular choice of parameters in the left panel. Note that the cross sections are largest for the case of negligible mixing angle and are considerably lower for a mixing angle close to the upper allowed value, $\theta_B = 0.3$. Therefore, we translate the experimental bounds on the cross section into bounds on the vev only for $\theta_B = 0$.  
The lower bounds on $v_B$ are strong in the region where $M_{h_B} < v_B$ and become weak only in the unphysical region where $ M_{h_B} \gg v_B$. To show the transition between these two regimes, we shade $M_{h_B} > v_B$ in gray in the plots. As one can appreciate these bounds are highly non-trivial, notice that the least constrained scenario is scenario I where the hypercharge of the new quarks are the same as the hypercharges of the SM quarks. In scenarios II and III the bounds are very strong and the symmetry breaking scale $v_B$ has to be well above the electroweak scale. 

\begin{figure}[tb]
\includegraphics[width=0.48\linewidth]{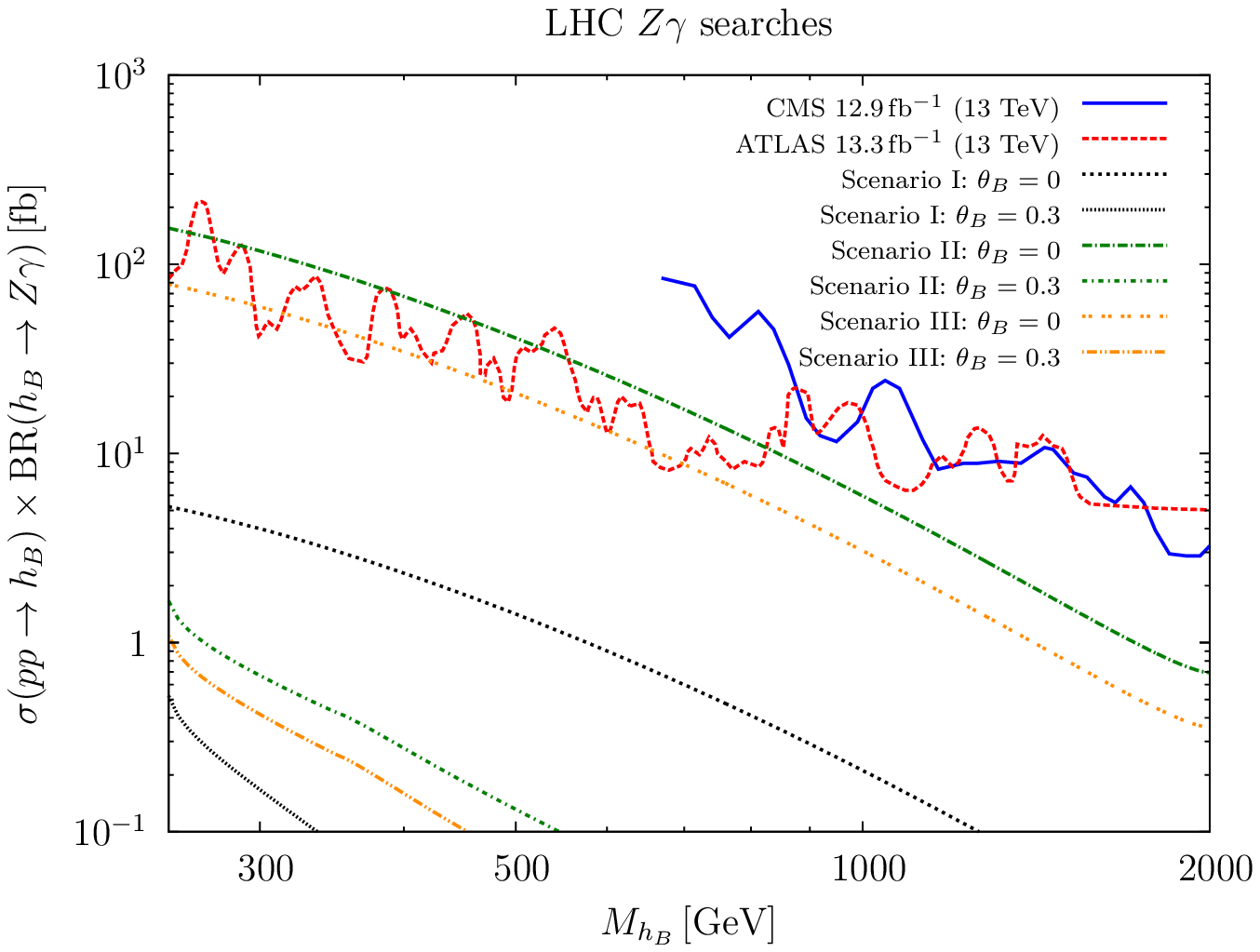}\hspace{.03\linewidth}
\includegraphics[width=0.48\linewidth]{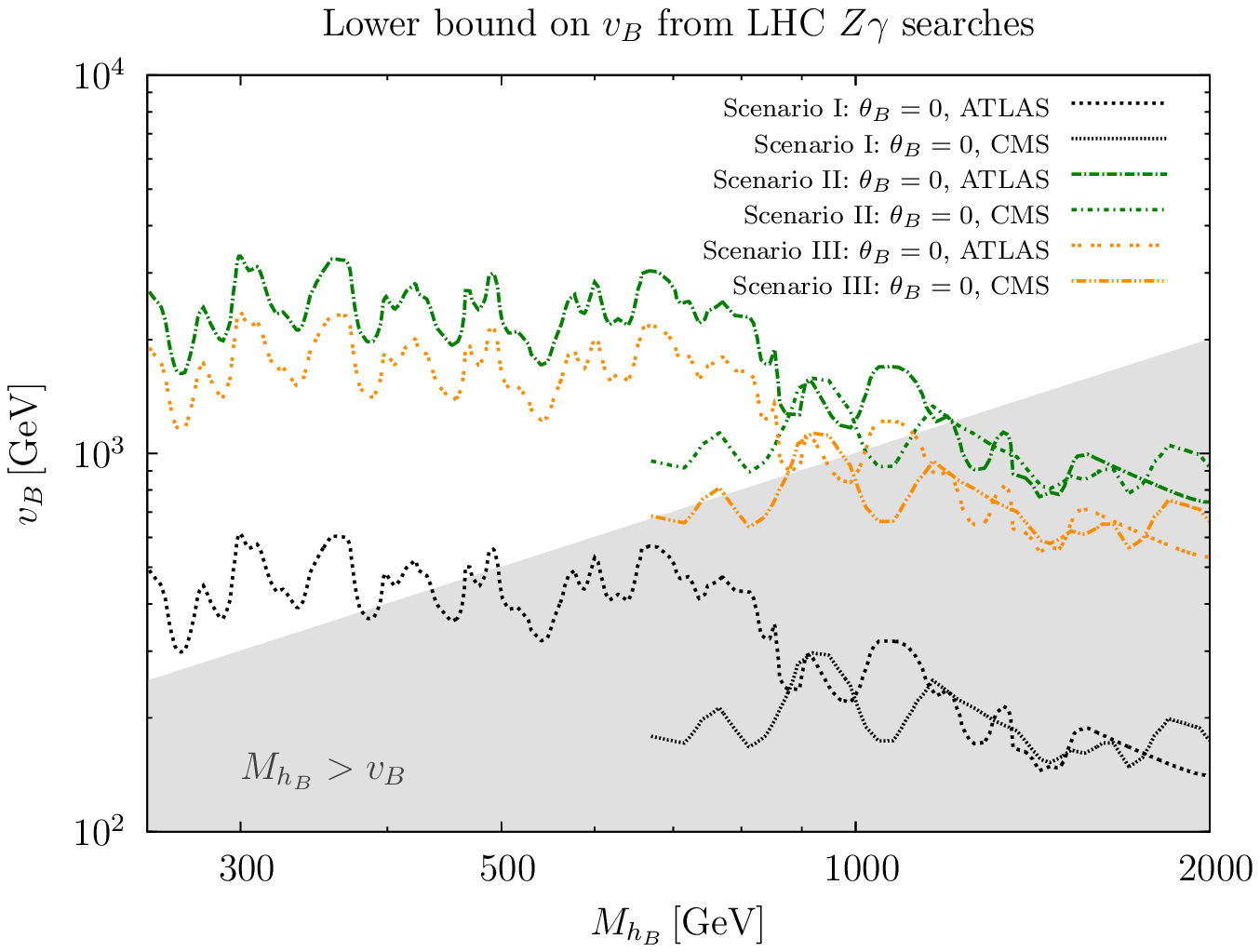}
\caption{Constraints from $Z \gamma$  searches~\cite{CMS:2016cbb,ATLAS:2016lri}.
Left panel: experimental bounds as well as cross section predictions in scenarios I, II, and II as a function of the baryonic Higgs mass. Cross sections are shown both for negligible mixing angle and for a near maximal value of $\theta_B = 0.3$. 
Right panel: experimental bounds translated into lower limits on $v_B$ as a function of $M_{h_B}$ for the case of negligible mixing. In the left panel we use $v_B=\unit[2]{TeV}$, and in both panels $n_f=3$ and $m_Q=\unit[1]{TeV}$.
\label{fig:ZgammaLimits}}
\end{figure}

\begin{figure}[tb]
\includegraphics[width=0.48\linewidth]{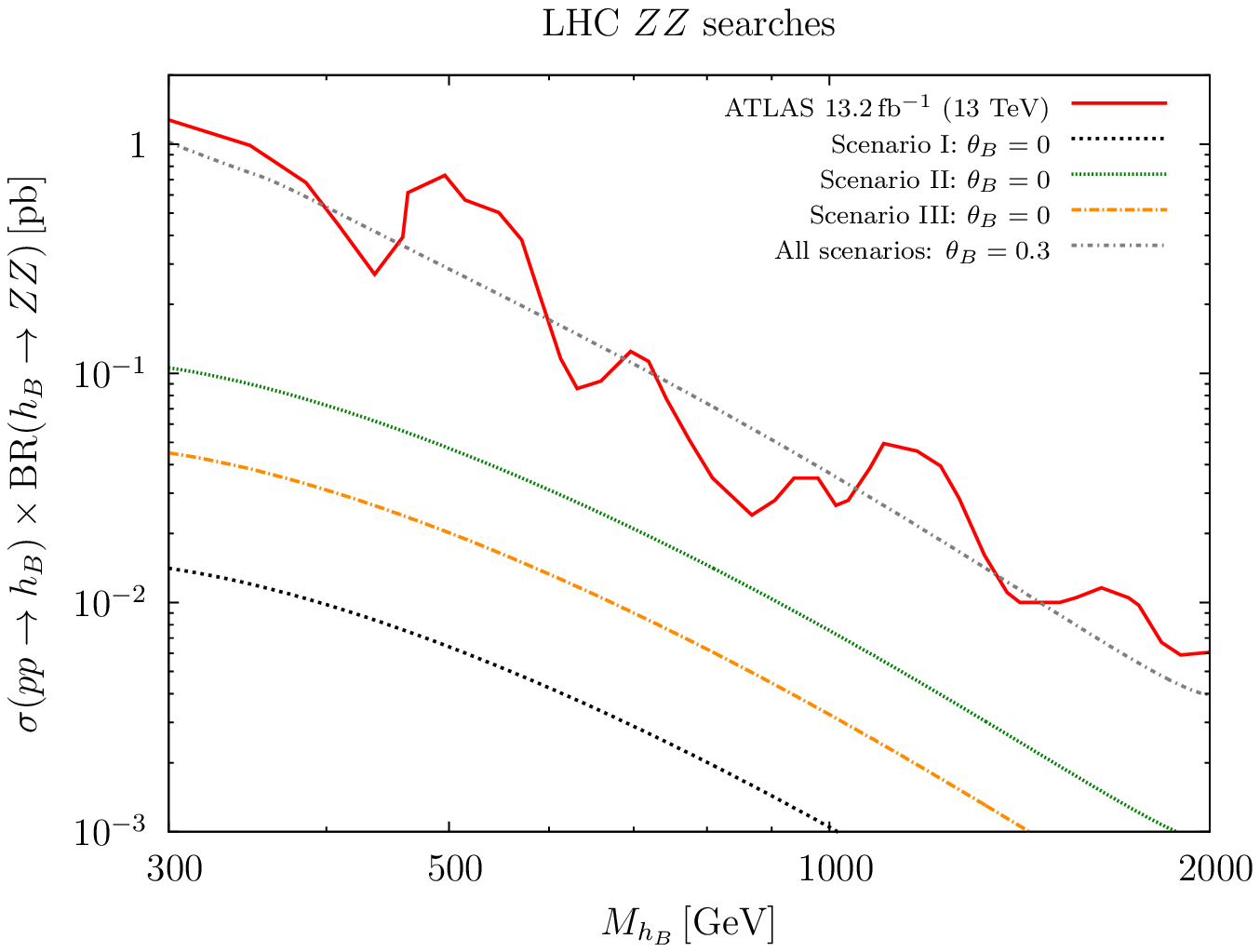}\hspace{.03\linewidth}
\includegraphics[width=0.48\linewidth]{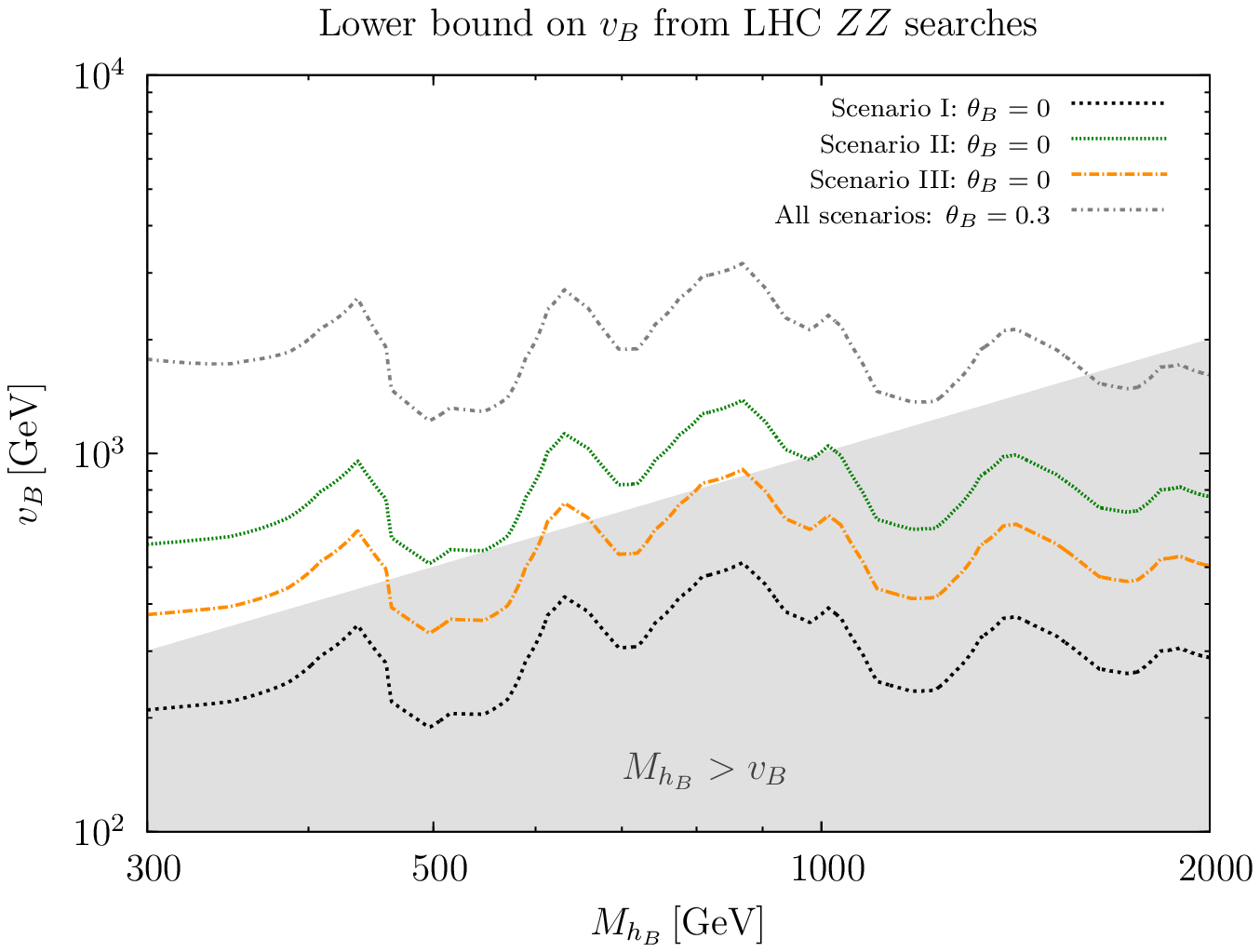}
\caption{Constraints from $Z Z$ searches~\cite{ATLAS:2016npe}.
Left panel: experimental bounds as well as cross section predictions in scenarios I, II, and II as a function of the baryonic Higgs mass. Cross sections are shown both for negligible mixing angle and for a near maximal value of $\theta_B = 0.3$ for which the three scenarios coincide. 
Right panel: experimental bounds translated into lower limits on $v_B$ as a function of $M_{h_B}$. In the left panel we use $v_B=\unit[2]{TeV}$, and in both panels $n_f=3$ and $m_Q=\unit[1]{TeV}$.
\label{fig:ZZLimits}}
\end{figure}

\begin{figure}[tb]
\includegraphics[width=0.48\linewidth]{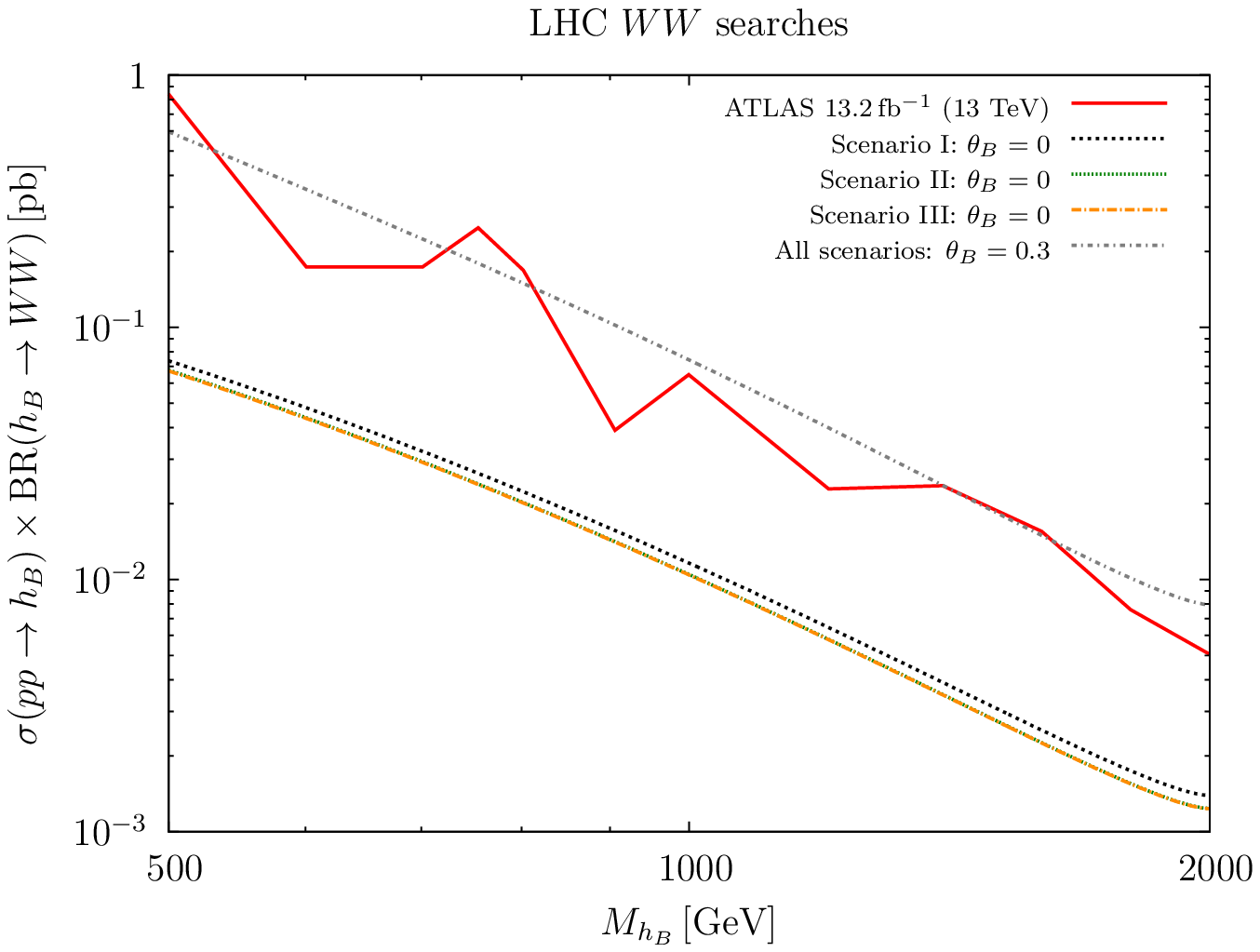}\hspace{.03\linewidth}
\includegraphics[width=0.48\linewidth]{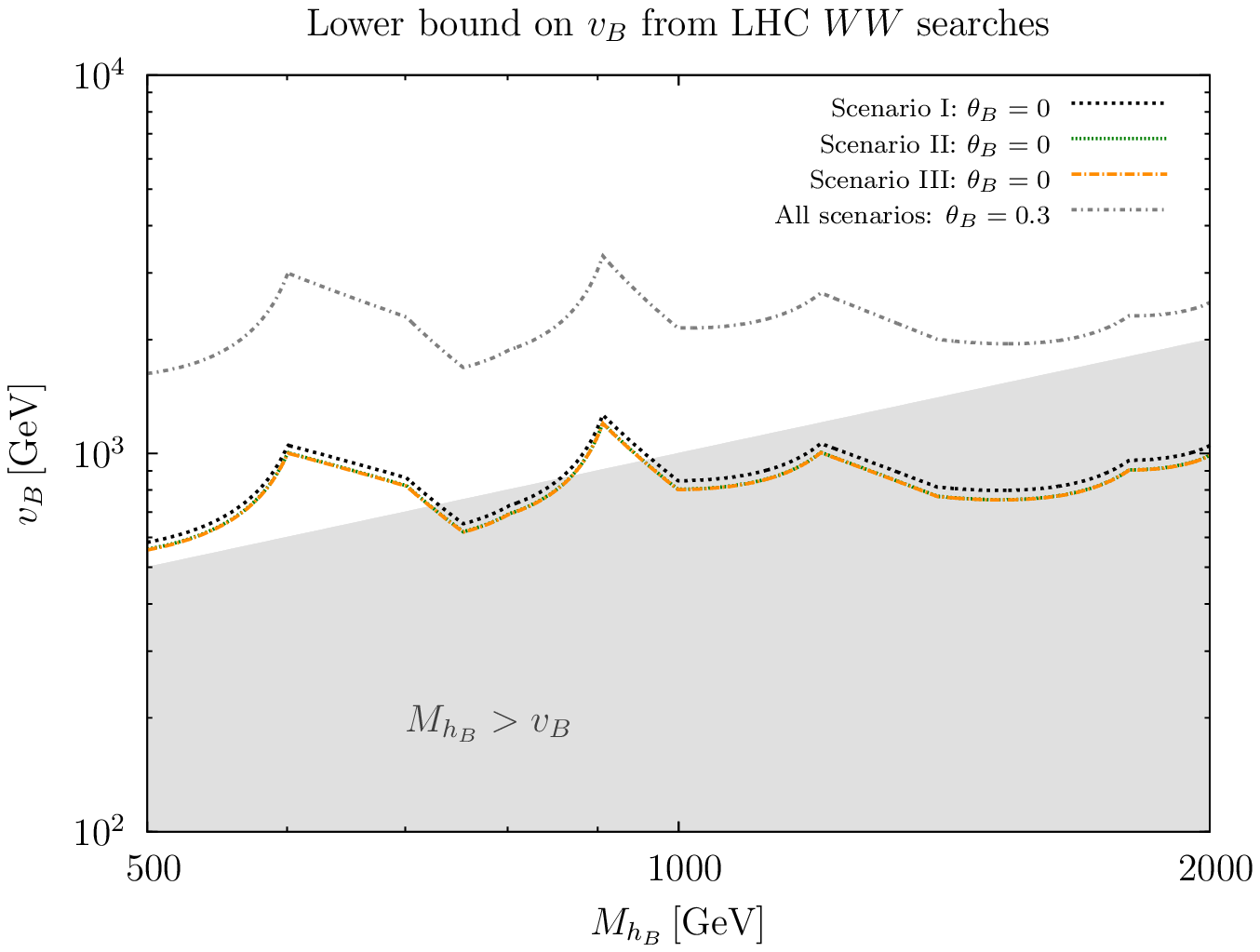}
\caption{Constraints from $WW$ searches~\cite{ATLAS:2016cwq}.
Left panel: experimental bounds as well as cross section predictions in scenarios I, II, and II as a function of the baryonic Higgs mass. Cross sections are shown both for negligible mixing angle and for a near maximal value of $\theta_B = 0.3$ for which the three scenarios coincide. 
Right panel: experimental bounds translated into lower limits on $v_B$ as a function of $M_{h_B}$. In the left panel we use $v_B=\unit[2]{TeV}$, and in both panels $n_f=3$ and $m_Q=\unit[1]{TeV}$.
\label{fig:WWLimits}}
\end{figure}

Other potentially relevant search channels are $Z\gamma$~\cite{CMS:2016cbb,ATLAS:2016lri}, $ZZ$~\cite{ATLAS:2016npe}, and $WW$~\cite{ATLAS:2016cwq}. Dijet searches turn out to be less constraining and will only be used for constraints on the leptophobic gauge boson in the next section. In Figures~\ref{fig:ZgammaLimits}, \ref{fig:ZZLimits}, and \ref{fig:WWLimits} we show the bounds from the $Z \gamma$, $ZZ$ and $WW$ channels, respectively. As for the $\gamma \gamma$ channel, we shade the region where $M_{h_B} > v_B$ in gray. Notice that the bounds from $WW$ and $ZZ$ become relevant for large mixing angles, where $\gamma \gamma$ and $Z \gamma$  are heavily suppressed. Since the tree-level decays to $WW$ and $ZZ$ dominantly proceed via mixing with the SM Higgs in the case of a large mixing angle, the cross sections and experimental bounds for the three scenarios coincide for $\theta_B = 0.3$

The result of the above discussion taking into account all current experimental bounds is that the symmetry breaking scale is highly constrained in most of the parameter space. For small mixing angle the most relevant bounds are from $\gamma \gamma$ searches and for large mixing angle the most relevant bounds are from $ZZ$ and $WW$ searches.

\subsection{Leptophobic Gauge Boson}
\label{sec:ZB}
These theories predict the existence of a leptophobic gauge boson. The most stringent constraints on this new gauge boson come from dijet searches. In Ref.~\cite{Fairbairn:2016iuf}, dijet constraints were tabulated in a model-independent way and we use these to constrain the leptophobic gauge boson, together with more recent experimental results~\cite{ATLAS:2016xiv,CMS:2017xrr,Aaboud:2017yvp}. In this section we neglect kinetic mixing between $U(1)_B$ and $U(1)_Y$. For a detailed discussion of kinetic mixing in these models see Ref.~\cite{Duerr:2014wra}.

\begin{figure}[tb]
\includegraphics[width=0.48\linewidth]{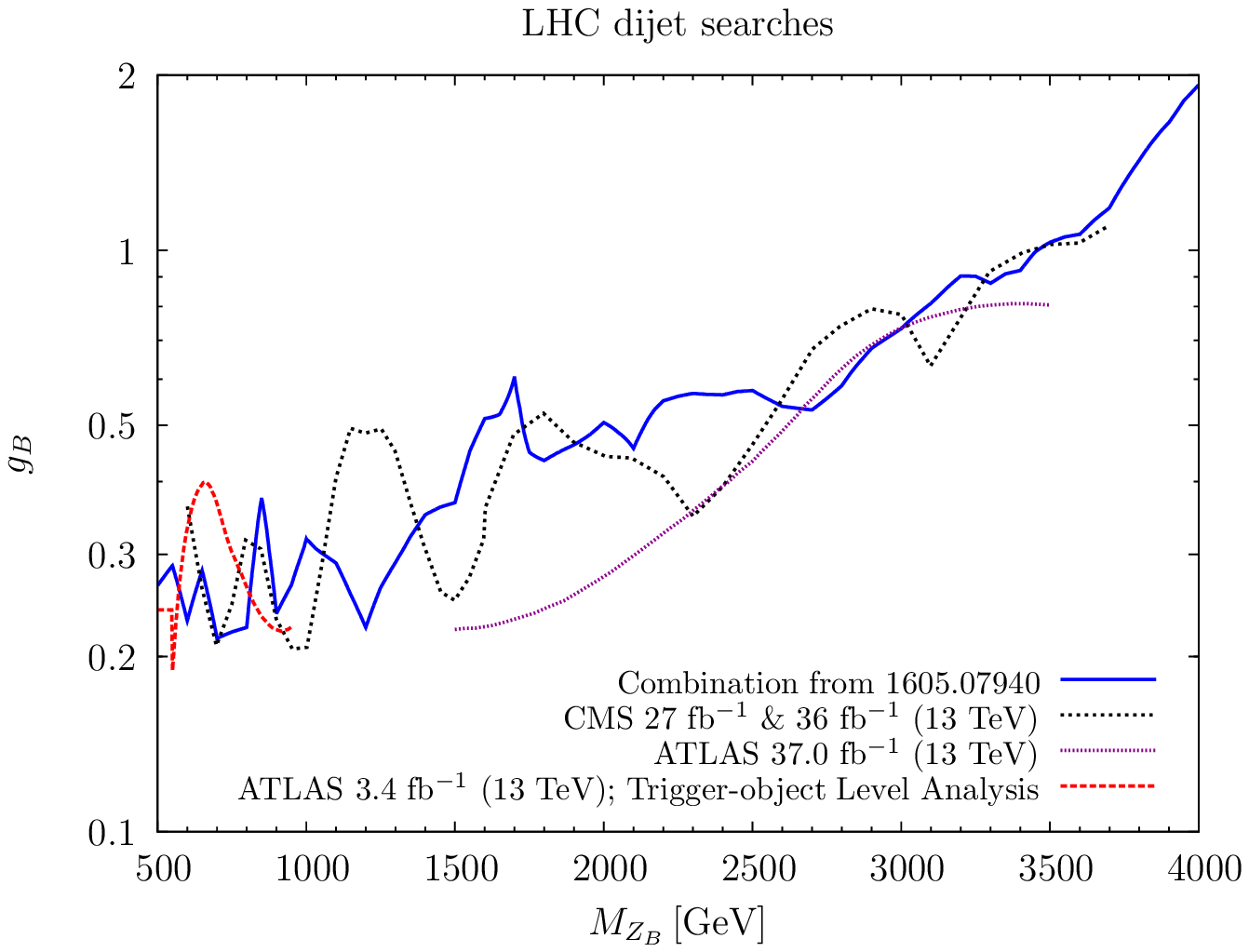}\hspace{.03\linewidth}
\includegraphics[width=0.48\linewidth]{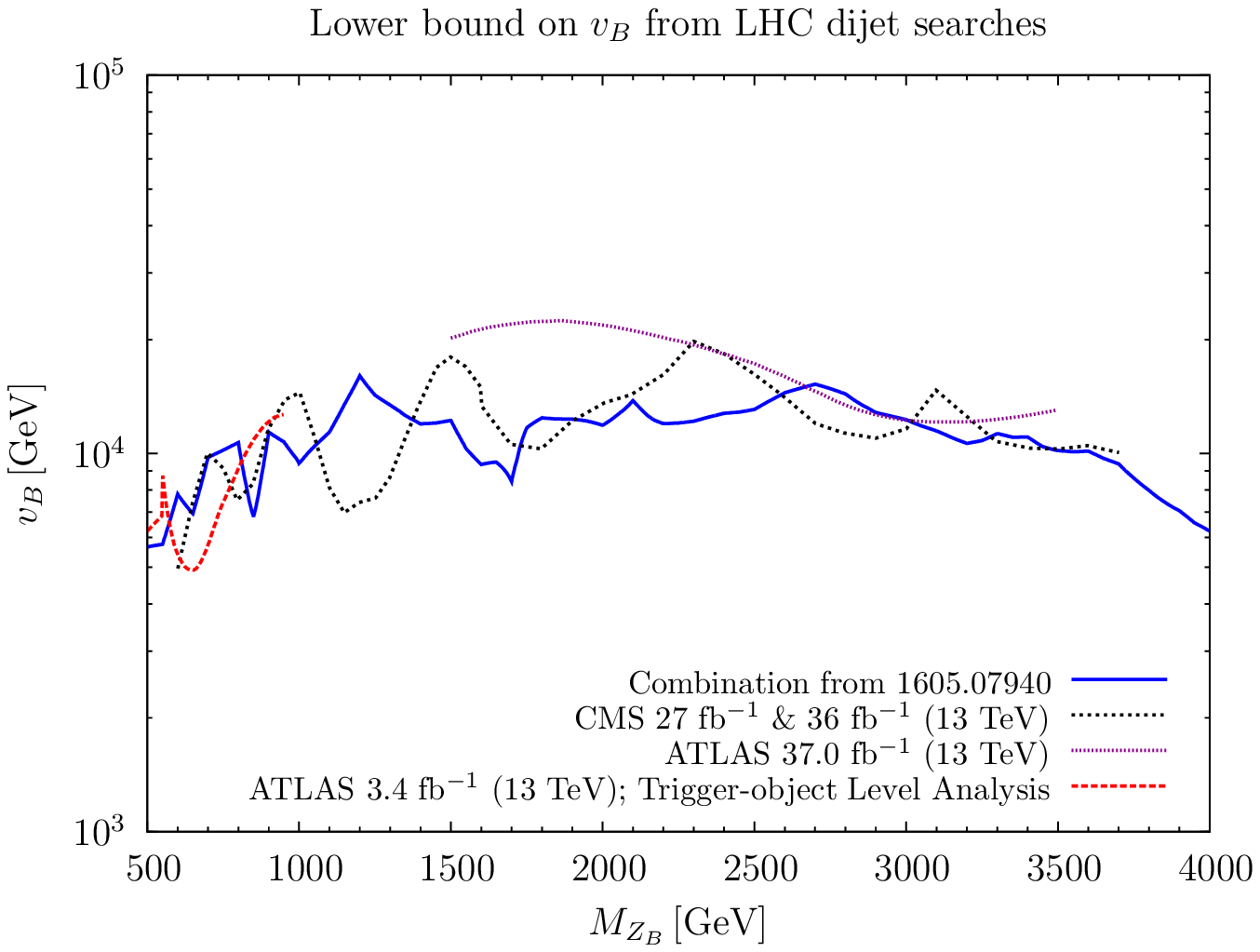}
\caption{Current experimental constraints on leptophobic gauge bosons from dijet searches~\cite{Fairbairn:2016iuf,ATLAS:2016xiv,CMS:2017xrr,Aaboud:2017yvp}. Left panel: upper bound on $g_B$ as a function of $M_{Z_B}$. Right panel: re-interpretation of the bounds in the mass--coupling plane as a lower bound on the symmetry breaking scale $v_B$.
\label{fig:dijetLimits}}
\end{figure}

In the left panel of Figure~\ref{fig:dijetLimits} we show the constraints in the $M_{Z_B}$--$g_B$ plane. These bounds assume that there are only decays into the SM quarks. If other decay channels are open, the bounds will be weaker. As we can see, the leptophobic gauge boson can be light when the gauge coupling $g_B$ is below 0.2. In the right panel of Figure~\ref{fig:dijetLimits} the gauge coupling limit is translated into a limit on the symmetry breaking scale. In combination with the need for a perturbative gauge coupling, $v_B$ cannot be smaller than $\unit[3.4]{TeV}$. We see that in the case of small scalar mixing the di-photon limits are competitive for all values of $M_{Z_B}$, compare with Figure~\ref{fig:diphotonLimits}. While for large scalar mixing the $WW$ and $ZZ$ channel are most sensitive to the symmetry breaking scale in the case of a heavy leptophobic gauge boson, compare to Figure~\ref{fig:ZZLimits}.

 \begin{figure}[bt]
\includegraphics[width=0.45\linewidth]{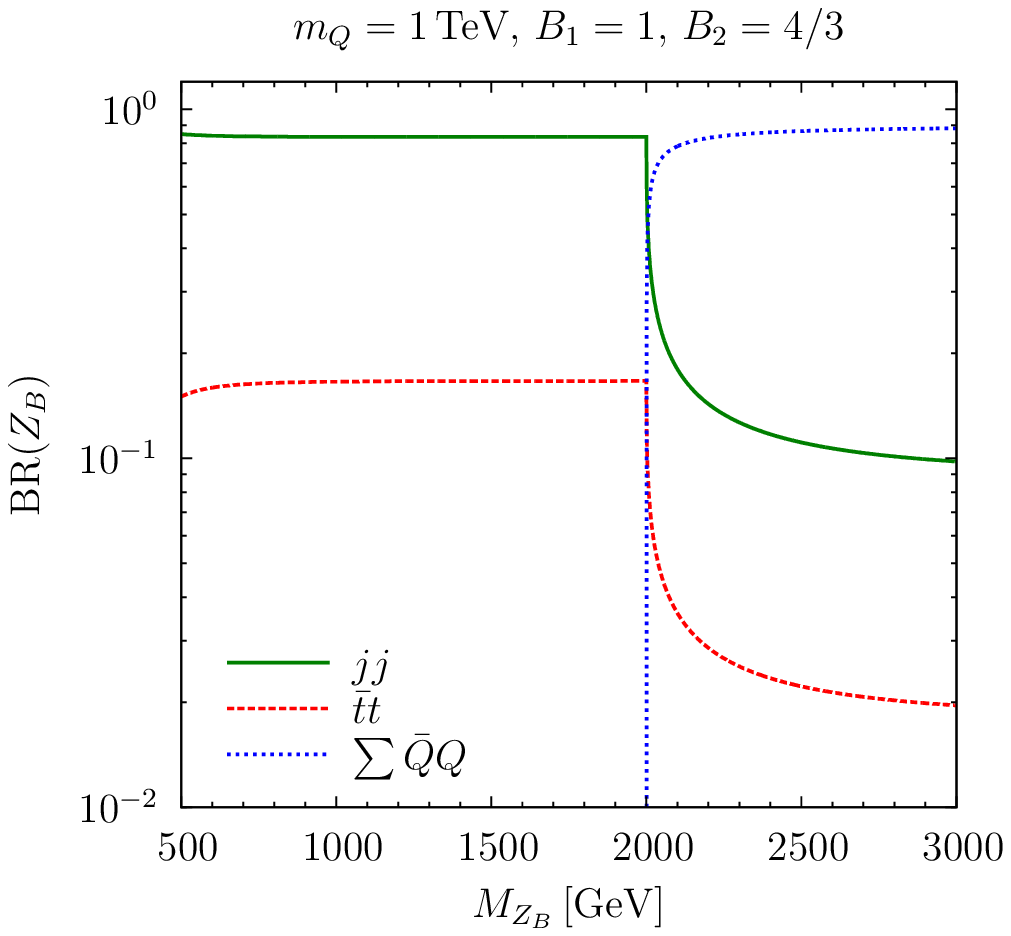}
\includegraphics[width=0.45\linewidth]{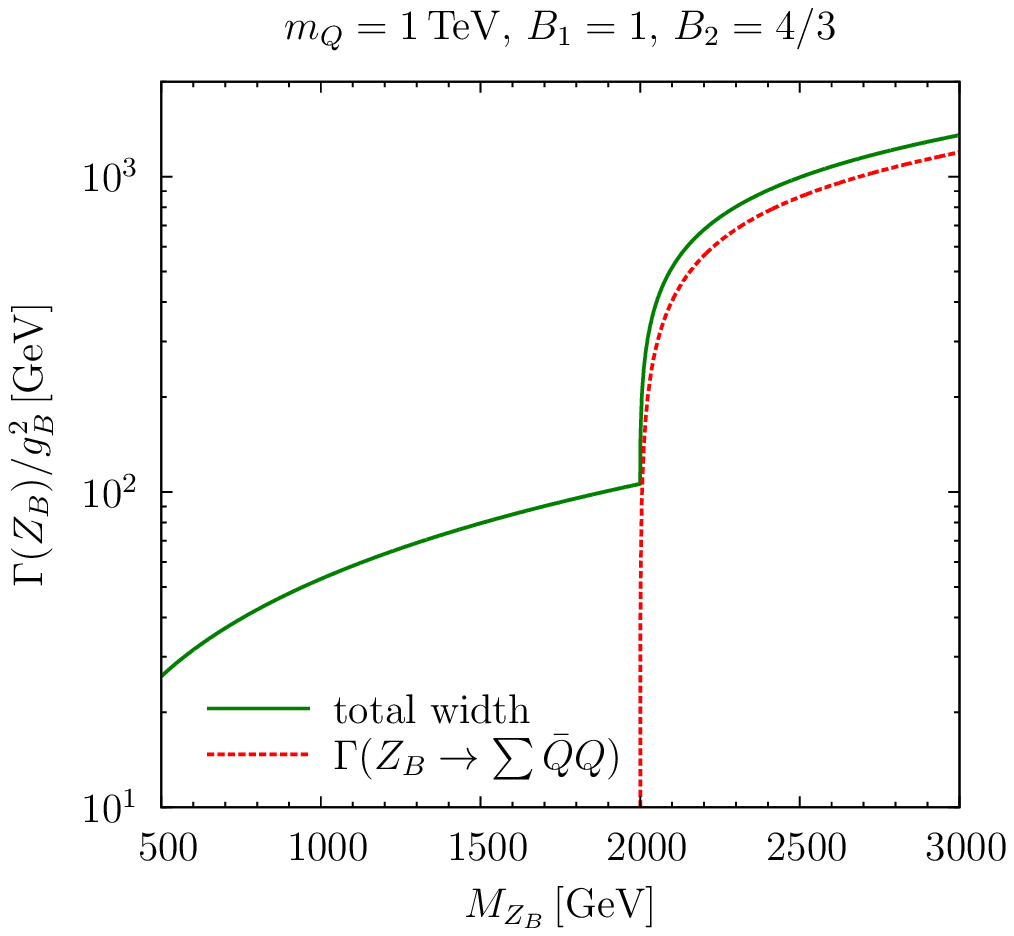}
\caption{Branching ratios and total decay width of the leptophobic gauge boson. Here we assume $m_Q=\unit[1]{TeV}$, $B_1=1$ and $B_2=4/3$ for illustration. 
The decays into all SM quarks except for the top quarks are included in the $jj$ decays.}
\label{fig:ZBBr}
\end{figure}

In order to understand the properties of the leptophobic gauge boson in these models we show in Figure~\ref{fig:ZBBr} the branching ratios and the total width. As one can appreciate the dominant decay channel is the decay into quarks when the decays into the vector-like quarks are not allowed kinematically. Once the decays into vector-like quarks are allowed they dominate since they have larger baryon number. Here we assume $B_1=1$ and $B_2=4/3$ for illustration. It is interesting to see that the total decay width is quite large for any gauge coupling $g_B$. The discovery of this 
leptophobic gauge boson is crucial for the testability of these theories.

\section{Summary}
\label{sec:Summary}
We have investigated the possible signatures at the Large Hadron Collider from decays of the new Higgs in simple extensions of the Standard Model where baryon number is a 
local symmetry spontaneously broken at the low scale. In this context one predicts the existence of a leptophobic gauge boson and a new CP-even 
Higgs boson associated to the spontaneous breaking of baryon number, as well as new vector-like quarks needed to cancel the baryonic anomalies. 

The baryonic Higgs $h_B$ can decay into all the SM particles, the leptophobic gauge boson and into the vector-like quarks, giving rise to peculiar signatures at the LHC. 
We have investigated the constraints from $\gamma \gamma$, $Z\gamma$, 
$ZZ$, and $WW$ searches at the LHC. We find scenarios where the branching ratio into two photons can be large and 
one can have very striking signatures at the LHC.  We also studied the properties of the 
leptophobic gauge boson present in these theories. We have shown that di-boson searches for the decays of the baryonic Higgs and di-jet searches for the leptophobic gauge boson 
are highly complementary and efficiently explore the parameter space of these theories.
We find a lower bound one the symmetry breaking scale using the constraints from the di-photon searches. 
The signatures studied in the article are crucial to understand the testability of these theories at the LHC.  

\section*{Acknowledgments}
P.F.P.\ thanks Mark B.\ Wise for discussions. The work of P. F. P. was supported by the U.S. Department of Energy under contract No.DE-SC0018005.
M.D.\ is supported by the German Science Foundation~(DFG) under the Collaborative Research Center~(SFB) 676 `Particles, Strings, and the early Universe' as well as the European Union's Horizon 2020 research and innovation programme under ERC starting grant agreement no.\ 638528 (`NewAve'). 
\appendix
\section{Feynman Rules}
\label{app:FeynmanRules}
\begin{align}
\overline{U_1} U_1 Z &: \ - \frac{i e}{\sin 2 \theta_W} \left( 1 - 2\,Q_u \sin^2 \theta_W \right) \gamma^\mu, \\
 \overline{U_2} U_2 Z &: \ i \,Q_u\, e \tan \theta_W  \ \gamma^\mu, \\
\overline{D_1} D_1 Z &: \ \frac{i e }{\sin 2 \theta_W} \left( 1 + 2 Q_d \sin^2 \theta_W \right) \ \gamma^\mu, \\
\overline{D_2} D_2 Z &: \ i Q_d \, e \tan \theta_W  \ \gamma^\mu, \\
\overline{U_i} U_i A &: \ - Q_u \, i e\, \gamma^\mu, \\
\overline{D_i} D_i A &: \ - Q_d \, i e \gamma^\mu, \\
\overline{U_k} U_k h_B&: \ i \frac{M_{U_k}}{v_B} \cos \theta_B, \\
\overline{D_k} D_k h_B&: \ i \frac{M_{D_k}}{v_B} \cos \theta_B, \\
h_1 h_1 h_B&: c_{11B} \equiv - 6 v_0 \lambda_H \cos^2 \theta_B \sin \theta_B + 6 v_B \lambda_B \cos \theta_B \sin^2 \theta_B \nonumber \\
           &\hphantom{:} + \lambda_{HB} \left( v_B \cos^3 \theta_B + 2 v_0 \cos^2 \theta_B \sin \theta_B - 2 v_B \cos \theta_B \sin^2 \theta_B - v_0 \sin^3 \theta_B \right).
\end{align}
\section{Decay Widths}
\label{app:DecayWidths}
\subsection{Baryonic Higgs}
The tree-level decay widths of the baryonic Higgs are given as follows. Here, $f$ denotes a SM fermion or a new quark. 
\begin{align}
 \Gamma (h_B \to \bar{f} f) &=\frac{N_c}{8 \pi} 
|c_{h_B \bar{f} f}|^2 
M_{h_B} \left( 1 \ - \ 4 \frac{M_f^2}{M_{h_B}^2} \right)^{3/2}, \\
 \Gamma (h_B \to h_1 h_1) &= \frac{1}{32 \pi}
 \frac{|c_{11B}|^2}{ M_{h_B}} 
\left( 1 - 4 \frac{M_{H_1}^2}{M_{h_B}^2} \right)^{1/2}, \\
 \Gamma (h_B \to W W) &= \frac{G_F}{8 \sqrt{2}\pi}  \sin^2 \theta_B \ M_{h_B}^3 \left( 1 \ - \ 4 \frac{M_W^2}{M_{h_B}^2} + 12 \frac{M_W^4}{M_{h_B}^4}\right) \left( 1 - 4 \frac{M_W^2}{M_{h_B}^2}\right)^{1/2} , \\
 \Gamma (h_B \to Z Z) &= \frac{G_F}{16 \sqrt{2}\pi}  \sin^2 \theta_B \ M_{h_B}^3 \left( 1 \ - \ 4 \frac{M_Z^2}{M_{h_B}^2} + 12 \frac{M_Z^4}{M_{h_B}^4}\right) \left( 1 - 4 \frac{M_Z^2}{M_{h_B}^2}\right)^{1/2} , \\
 \Gamma (h_B \to Z_B Z_B) &= \frac{1}{32 \pi}  \frac{\cos^2 \theta_B \ M_{h_B}^3}{v_B^2} \left( 1 \ - \ 4 \frac{M_{Z_B}^2}{M_{h_B}^2} + 12 \frac{M_{Z_B}^4}{M_{h_B}^4}\right) \left( 1 - 4 \frac{M_{Z_B}^2}{M_{h_B}^2}\right)^{1/2}.
\end{align}
$c_{11B}$ is defined in the Feynman rules in Appendix.~\ref{app:FeynmanRules}. $c_{h_B \bar{f} f}$ is the coupling of the baryonic Higgs to the fermion $f$, which is given by
\begin{equation}
c_{h_B \bar{f} f} = \frac{M_f}{v_0} \sin \theta_B
\end{equation}
for SM fermions and  
\begin{equation}
c_{h_B \bar{f} f} = \frac{M_f}{v_B} \cos \theta_B
\end{equation}
for the new vector-like quarks. 

The loop-induced partial decay widths to SM gauge bosons of the new Higgs boson $h_B$ under the assumption that there is only negligible mixing with the SM Higgs are given in Appendix~B of Ref.~\cite{Duerr:2016eme} and will not be repeated here. We used \texttt{Package-X}~\cite{Patel:2015tea,Patel:2016fam} for the calculation of one-loop integrals. If there is non-negligible mixing with the SM Higgs, also SM particles will run in the loops, and corresponding formulas can for example be found in Ref.~\cite{Djouadi:2005gi}.
\subsection{Leptophobic gauge boson}
The partial decay width of the leptophobic gauge boson $Z_B$ to SM quarks is given by
 \begin{equation}
  \Gamma ( Z_B \rightarrow \bar{q} q ) = \frac{g_B^2}{36 \pi} M_{Z_B} \left(1 - \frac{4 M_q^2}{M_{Z_B}^2} \right)^{\frac{1}{2}} \left(1 + \frac{2 M_q^2}{M_{Z_B}^2} \right),
 \end{equation}
while the decay width to vector-like quarks $Q$ with mass $m_Q$ is given by
\begin{equation}
 \Gamma ( Z_B \rightarrow \bar{Q}Q ) =\frac{g_B^2}{8 \pi} M_{Z_B} \left(1 - \frac{4 m_Q^2}{M_{Z_B}^2} \right)^{\frac{1}{2}} \left[ \left( B_1^2 + B_2^2 \right) \left(1-\frac{m_Q^2}{M_{Z_B}^2}\right) + 6 B_1 B_2 \frac{m_Q^2}{M_{Z_B}^2} \right].
\end{equation}



\end{document}